\documentstyle[aaspp4,12pt]{article}

\def\om{\Omega_M}
\def\ol{\Omega_\Lambda}
\def\sinn{\mathop{\rm sinn}\nolimits}
       
\newcommand{\be}{\begin{equation}}
\newcommand{\ee}{\end{equation}}
\newcommand{\bea}{\begin{eqnarray}}
\newcommand{\eea}{\end{eqnarray}}
\def\ceqn(#1){equation~(\ref{#1})}
\def\ceq(#1){(\ref{#1})}
\def\eqn(#1){\label{#1}}

\newcommand{\MLCS}{{\rm MLCS}}
\newcommand{\TF}{{\rm TF}}
\newcommand{\SF}{{\rm SF}}

\newcommand{\OM}{\Omega_M}
\newcommand{\OL}{\Omega_\Lambda}

\newcommand{\cpar}{{\cal C}}
\newcommand{\spar}{{\cal S}}

\newcommand{\like}{{\cal L}}

\newcommand{\eps}{\epsilon}

\begin{document}
\title{Type Ia Supernovae, Evolution, and the Cosmological Constant}

\author{Persis S. Drell,\altaffilmark{1}  Thomas J. Loredo,\altaffilmark{2} and Ira Wasserman\altaffilmark{1,2}}
\altaffiltext{1}{Floyd R.\ Newman Laboratory of Nuclear Studies,
Cornell University, Ithaca, NY 14853-5001}
\altaffiltext{2}{Center for Radiophysics and Space Research,
Cornell University, Ithaca, NY 14853-6801}

\begin{abstract}
We explore the possible role of evolution in the analysis of data on
SNe Ia at cosmological distances. First, using a variety of simple
sleuthing techniques, we find evidence that the properties of the high
and low redshift SNe Ia observed so far differ from one another.  Next,
we examine the effects of allowing for an uncertain amount of
evolution in the analysis, using two simple phenomenological models for
evolution and prior probabilities that express a preference for no evolution
but allow it to be present. One model
shifts the magnitudes of the high redshift SNe Ia relative
to the low redshift SNe Ia by a fixed amount.  A second,
more realistic, model introduces a continuous
magnitude shift of the form 
$\delta m(z) = \beta \ln(1+z)$ to the SNe Ia sample.
The result is that cosmological models and
evolution are highly degenerate with one another, so that the
incorporation of even very simple models for evolution makes it
virtually impossible to pin down the values of $\Omega_M$ and
$\Omega_\Lambda$, the density parameters for nonrelativistic matter and
for the cosmological constant, respectively.  
The Hubble constant, $H_0$, is unaffected by
evolution.  We evaluate the Bayes
factor for models with evolution versus models without evolution,
which, if one has no prior predilection for or against evolution, is
the odds ratio for these two classes of models.  The resulting values
are always of order one, in spite of the fact that the models that
include evolution have additional parameters; thus, the data alone
cannot discriminate between the two possibilities.  Simulations show
that simply acquiring more data of the same type as are available now
will not alleviate the difficulty of separating evolution from
cosmology in the analysis.  What is needed is a better physical
understanding of the SN Ia process, and the connections among the
maximum luminosity, rate of decline, spectra,  and initial conditions, so that
physical models for evolution may be constructed, and confronted with
the data.  Moreover, we show that if SNe Ia evolve with time, but
evolution is neglected in analyzing data, then, given enough SNe Ia,
the analysis hones in on values of $\Omega_M$ and $\Omega_\Lambda$
which are incorrect.  Using Bayesian methods, we show that the
probability that the cosmological constant is nonzero (rather than
zero) is unchanged by the SNe Ia data when one accounts for the
possibility of evolution, provided that we do not discriminate among
open, closed and flat cosmologies a priori. The case for nonzero
cosmological constant is stronger if the Universe is presumed to be
flat, but still depends sensitively on the degree to which the peak
luminosities of SNe Ia evolve as a function of redshift.
\end{abstract}


\section{Introduction}

The realization that the rates of decline of the brightnesses of Type
Ia supernovae (SNe Ia) are correlated with their peak luminosities
(Phillips 1992) has led to renewed efforts to use them as cosmological
distance markers (Hamuy et al. 1995, 1996a, 1996b; Riess, Press and
Kirshner 1995, 1996). Ongoing searches for high redshift ($z\sim
0.5-1$) SNe Ia have employed phenomenological models for these
correlations to constrain the variation of luminosity distance $D_L(z)$
with redshift; the results have been interpreted to imply the existence
of a nonzero cosmological constant (Perlmutter et al.\ 1998 hereafter
P98, Riess et al.\ 1998 hereafter R98).  Moreover, the results appear
to rule out the simplest version of a flat cosmology, in which the
density parameter for ``ordinary'' matter (including as yet
unidentified nonbaryonic material) $\Omega_M=1$ and the density
parameter for the cosmological constant $\Omega_\Lambda=0$.

Although the logical possibility that $\Omega_\Lambda\sim 1$
today has long been recognized (Einstein 1917), it is anathema
to many theorists, since the associated vacuum energy density
must be $\rho_{vac}\sim 10^{-122}M_{\rm P}^4$, where $M_{\rm P}$
is the Planck mass. (Theoretical and conceptual problems
with a nonzero cosmological constant so small compared with
its ``natural'' scale have been reviewed by Weinberg 1989
and Carroll, Press \& Turner 1992.) On the other hand,
there is some evidence from large scale structure simulations
that $\Omega_\Lambda\neq 0$ in a flat Universe fits the
observations well (Cen 1998). Conceivably, what
we interpret as a cosmological ``constant'' might be an 
evolving field (e.g., Caldwell, Dave \& Steinhardt 1998;
Garnavich et al.\ 1998; Perlmutter, Turner \& White 1999). 
In any case, a convincing demonstration that the expansion
rate of the Universe is increasing would have a revolutionary
impact on our understanding of fundamental physics.

In view of the importance of the  potential discovery of a nonzero
cosmological constant, we have undertaken an independent study of the
published data in an effort to understand their implications better.
Our motivations are both phenomenological and astrophysical (and may
end up being related ultimately). On the phenomenological side, we note
that three different analysis methods are used to compute distances,
the multicolor light curve shape (MLCS) method (Riess, Press \&
Kirshner 1995, 1996), the M15 or template fitting (TF) method (Phillips
1992; Hamuy et al.\ 1995, 1996a, 1996b),
and the stretch factor (SF) method (Perlmutter et al.\ 1997).  None of
these methods is a perfect description of reality. As we will show,
they are not always in agreement,  and there seems to be no physical or
phenomenological reason to prefer one to the other.

On the astrophysical side, we note that there are processes such
as evolution of the supernovae sample
that can mimic the effects of cosmology at high redshifts and which
are extremely difficult to constrain convincingly
with the current data.
Therefore, it is useful to ask at what level the current data
are able to distinguish the effects of cosmology from these
other processes. 
We find that allowing for the possibility of 
a redshift-dependent shift in SNe Ia peak magnitudes
(such that the most distant observed SNe Ia are dimmed by $\approx 0.2$
to 0.3 mag) renders $\OM=1$ and $\OL=0$ acceptable, and that this
is true for a variety of phenomenological models for the evolution.
We also present results of simulations that show
that if SNe Ia luminosities evolve with redshift, but evolution
is neglected in analyzing the data, then, given enough data,
the analyses will settle on precisely determined, but incorrect,
values for $\Omega_M$ and $\Omega_\Lambda$, and that the
incorrectness of the model will not be detectable with a
standard $\chi^2$ goodness-of-fit test.  However, we find that
the Hubble constant, $H_0$, is virtually unaffected by evolution.

We believe that it is unjustifiable to try to determine
cosmological parameters $\OM$ and $\OL$ from data on ``standardized''
candles, such as the peak luminosities of SNe Ia, without 
allowing for the possibility of
source evolution. Our attitude is that an uncertain amount of
evolution must be presumed to occur, as a default; and the sensitivity
of the results to the uncertainty must be studied. 
Hopefully, one can demonstrate
from the data that source evolution is absent or negligible. 
\footnote{The observers have employed supplementary measurements---such as
source spectra---to argue that there is no compelling evidence
{\it for} evolution in the SNe Ia samples. If the physical
connection between the additional data and the estimators
of peak luminosity can be understood quantitatively, then
the supplementary measurements can be incorporated usefully
into analyses of models with evolution. Without
such physical understanding, these additional data cannot be
marshalled to argue {\it against} evolution.}
If that turns out to be untrue, as recent 
examination of the risetimes of the light curves
of the SNe Ia sample has preliminarily indicated
(Riess, Filippenko, Li, \& Schmidt 1999),
one might hope instead to constrain
the parameters in an evolutionary model along with the cosmological
parameters. Optimistically, one would anticipate that this might
be accomplished once enough data are acquired. We argue, from
simulations employing simple, phenomenological models, that such 
optimism may be unrealistic. What is needed is a better
physical understanding of the SN Ia process and its evolution
with redshift, before cosmological parameters can be determined
reliably from SNe Ia catalogues.  Such an understanding is
currently being sought by theorists (see, e.g., von Hippel, Bothun,
\& Schommer 1997; H\"oflich, Wheeler,
\& Thielemann 1998; Dom\'inguez et al.\ 1999).

In using observations of SNe Ia to determine cosmological parameters,
the raw data are combined by any of the three methods mentioned above
to derive single parameter summaries -- the distance moduli to the
sources. When a single catalogue of data is subjected to different
types of analysis, each of which derives one quantity per source, the
results of the individual analyses need not agree with one another
entirely, and there is information contained in the degree to which the
answers derived by the different methods differ\footnote{A simple
example might involve computing mean values of data using similar, but
not identical, weighting functions. Each weighted mean would then be a
slightly different superposition of all of the moments of the data
computed with one fiducial weighting function.}.  In the example under
consideration, the different analysis methods might probe slightly
different physical aspects of the SN Ia mechanism, and their
relationships to reality and to one another may be different at high
and low redshift.  Indeed, one clue that evolutionary effects are
important would be a systematic drift with increasing redshift between
the distance moduli implied by the MLCS and TF methods for the SNe Ia
observed and analyzed by R98 where identical SNe Ia data are subjected
to two different analysis methods.

In order to set the scale of interest for our investigations of
potential systematic effects in these data we plot, in Figure~1, the
joint credible regions for $\Omega_M$ and $\Omega_\Lambda$ for the
largest available data set (P98), analyzed as published (Figure~1$a$)
and after introducing a systematic offset to all the high redshift
distance moduli ($z > 0.15$) of $-0.1$ magnitudes (Figure 1$b$).  We
see that a correlated systematic shift of this size would have a major
impact on the interpretation of the data.  A nonzero cosmological
constant would still be favored, but the statistical significance of
the result would be much reduced.

In Section \ref{sec:review} we review the published data
that we employ in our study, as well as the salient features
of the light curve fitting methods. In Section \ref{sec:fit}
we compare the different fitting methods on a supernova by 
supernova basis where possible.  In Section \ref{sec:stat} we
explore whether the data have sufficient shape information to
distinguish effects of cosmology from other cosmological effects such 
as evolution.  We summarize our conclusions in Section 
\ref{sec:conc}.

\section{Measurements of $\om$ and $\ol$ Using Type Ia Supernovae }
\label{sec:review}

The traditional measure of distance to a SN is its distance modulus,
$\mu\equiv m_{\rm bol}-M_{\rm bol}$, the difference between its
bolometric apparent magnitude, $m_{\rm bol}$, and its bolometric absolute
magnitude, $M_{\rm bol}$.
In the Friedman-Robertson-Walker (FRW) cosmology,
when the (relative) peculiar velocity of
the source is negligible, the distance modulus is determined
by the source's redshift, $z$, according to
\begin{eqnarray}
\mu &=& 5 \log \left[D_L(z;\Omega_M,\Omega_\Lambda,H_0)\over 
      1\;\hbox{Mpc}\right] +25 \nonumber\\
  &\equiv& f(z; \Omega_M,\Omega_\Lambda,H_0).
\label{eq:mu} 
\end{eqnarray}
Here the luminosity distance $D_L(z;\Omega_M,\Omega_\Lambda,H_0)
=cH_0^{-1}d_L(z;\Omega_M,\Omega_\Lambda)$, where $c$ is the velocity of
light, $H_0$ is Hubble's constant at the present epoch, and the
dimensionless luminosity distance from redshift $z$ is
\begin{equation}
d_L(z;\Omega_M,\Omega_\Lambda)= 
(1+z)|\Omega_k|^{-1/2} \sinn \{ |\Omega_k|^{1/2} 
\int_0^z dz [ (1+z)^2(1+\om z) - z(2+z)\ol ]^{-1/2}\}, 
\end{equation}
with $\Omega_k = 1-\om-\ol$, and $\sinn(x)=\sinh(x)$ for $\Omega_k \geq 0$
and $\sin(x)$ for $\Omega_k \leq 0$~(e.g., Carroll, Press \& Turner 1992).
In principle, one could infer the cosmological parameters
$H_0$, $\OM$, and $\OL$ from the distribution of measured distance moduli
of sources at a variety of redshifts.

Several factors complicate implementation of such an analysis.  In
reality, bolometric data are not available, and one must infer $\mu$
using magnitudes $m_X$ and $M_X$ in some bandpass, $X$.  The bandpass
maps to a different region of the spectrum as a function of redshift
$z$, so $\mu$ cannot be calculated simply by taking the difference
between band-limited magnitudes; a $K$-correction term must be added
whose value depends not only on the source's redshift, but also on its
spectrum.  In addition, extinction along the line of sight increases
the apparent magnitude by some amount $A_X$ not due to the cosmological
effects modelled in \ceqn(eq:mu).  Further, the absolute magnitude of
the source---bolometric or band-limited---is not directly measured, but
must instead be inferred from other source properties.  Finally, the
inevitable presence of statistical uncertainties and peculiar
velocities further complicates straightforward use of \ceqn(eq:mu).

Of these complications, the need to infer the absolute
magnitude indirectly is the most troublesome.  Ideally, one seeks a population
of ``standard candles'' such that all members of the population
have the same $M$ (for convenience we henceforth drop the subscript
$X$).  If such a population could be identified,
the parameters $\OM$ and $\OL$ could be inferred even if the actual
value of $M$ for the population
were unknown (the remaining parameter, $H_0$, would remain
undetermined).  Historically, all attempts to identify such a
population have failed.  Particularly worrisome is the possibility that
some classes of objects that appear to be approximately
standard candles locally (at low redshift, where they can be
studied in detail) have evolved significantly, so that their
younger counterparts at high redshifts have different absolute
magnitudes, thwarting their use as cosmological distance indicators.

SNe Ia were briefly considered promising candidates for standard candles,
but observers quickly discovered that SNe Ia are not all identically
bright (Branch, 1987; Barbon, Rosino, \& Iijima 1989; Phillips et al.\
1987, 1992; Filippenko, et al.\ 1992a, 1992b; Leibundgut, et al.\ 1993).
The intrinsic dispersion in the peak absolute magnitudes of SNe Ia,
determined from studies of nearby events, is approximately 0.3 - 0.5 mag
(Schmidt et al.\ 1998). However, there is an apparent empirical
correlation between the rate of decline of the light curve of a given
SN Ia and its luminosity at maximum brightness that was first
quantified by Phillips (1992).  Various techniques have been developed to
take advantage of this correlation to determine the absolute magnitudes
of individual supernovae using their light curves (Phillips 1992; Hamuy
et al.\ 1995, 1996a, 1996b; Riess, Press \& Kirshner 1995, 1996;
Perlmutter et al.\ 1997); the relationships used in these analyses have
come to be known generically as ``Phillips relations.''  When applied to
nearby SNe Ia, these methods reduce the dispersion of the distance
moduli about the low-$z$ FRW distance modulus vs.\ redshift relation to
$\approx 0.15$ (Hamuy et al. 1996a, Riess, Press \& Kirshner 1996).

The goal of the high redshift supernovae searches is to observe a large
sample of supernovae at relatively large $z$, and understand their properties
well enough to infer reliable distance moduli for them, allowing
accurate determination of cosmological parameters.  Two experimental
groups have recently announced and published results from their
independent programs to discover and study high redshift supernovae for
this purpose (Perlmutter 1997 and P98; R98).  The resulting two data
sets share many low redshift SNe discovered by previous surveys, but
include different high redshift SNe, and differ in their analysis
methods.  We take advantage of the similarities and differences among
the data and methods used to assess the consistency or inconsistency of
the assumptions underlying the analyses.

P98 have published data on 60 SNe Ia.  Of these, 18 were discovered and
measured in the Cal\'an-Tololo survey (all at low redshift; Hamuy et
al.\ 1996c), and this group discovered 42 new SNe Ia at redshifts
between 0.17 and 0.83.  The $\mu$ values are inferred using the SF
light curve fitting method and are typically uncertain to $\pm 0.2$
magnitudes (``$1\sigma$'').  The SF method (Perlmutter et al.\ 1997;
P98) is based on fitting a time-stretched version of a single standard
template to the observed light curves.  The stretch factor, $s$, is
then used to estimate the absolute magnitude of the SNe Ia via a linear
relationship that is determined jointly with the cosmological
parameters.  The quoted $\mu$ values include a correction for
extinction in the Galaxy based on the detailed model of Burstein \&
Heiles (1982).

R98 have published results based on 50 SNe Ia. Of these, 37, including
27 at low redshift ($z<0.15$) and ten at high redshift ($z>0.15$) have
well-sampled light curves in addition to spectroscopic information; the
quoted ``$1\sigma$'' uncertainties for $\mu$ for these SNe Ia are
typically smaller than $\pm 0.2$ mag at high $z$ for determinations by
either MLCS or TF light curve fitting method.  The data for 17 of the
SNe Ia at low redshift come from the Cal\'an-Tololo survey (Hamuy et
al. 1996c).  We focus our attention on these 37 best-observed SNe Ia,
which dominate the analysis in R98.  These authors extensively
tabulate their reduced data and provide detailed information
about their fitting techniques, thus facilitating independent analysis
of their conclusions.

R98 employ two different methods to estimate the distance modulus based
on information from the light curves.  The TF method~(Hamuy et al.
1996a) fits a set of light curve templates with different values of
$\Delta m_{15}$, the total decline in brightness from peak to 15 days
afterward, to observations of a particular SN Ia. By interpolating
between the values of $\chi^2$ for the fits to the various templates, a
minimum $\chi^2$ value of $\Delta m_{15}$ for the SN Ia is estimated.
The peak absolute magnitude is deduced from the independently
calibrated linear relationship between $M$ and $\Delta m_{15}$. The
MLCS method consists of fitting an observed light curve to a
superposition of a standard light curve and weighted additional
templates that parametrize the differences among SNe Ia (e.g., Riess,
Press \& Kirshner 1996); the outcome of the fits for a particular SN Ia
consists of the weights associated with the deviations from the
standard, which in turn determine the difference between its peak
absolute magnitude and the standard's. The fits are done for more than
one color, and reddening and extinction are inferred from color
dependences (Riess, Press \& Kirshner 1996, R98).  Originally the MLCS
method used a rather small training set to determine the requisite
templates (Riess, Press \& Kirshner 1996), but R98 now train on a
considerably larger set of nearby SNe Ia to find them.  Both the MLCS
and TF methods are calibrated on nearby SNe Ia in the Hubble flow 
$(z < 0.15)$ and then applied to the SNe Ia discovered at high redshift.
The quoted $\mu$ values include a correction for local extinction
derived from the Burstein and Heiles model, and in addition the MLCS
method uses color dependence to estimate corrections for the extinction
and reddening due to absorbing material in the host galaxy.

Schematically, we can consider a lightcurve fitting method to
estimate the distance modulus for supernova number $i$ according to
the following model:
\be
\mu_i = m_i - (M_0 + \Delta_i).
\eqn(mu-model)
\ee
Here $m_i$ is the peak apparent magnitude for the SN, $M_0$ is
a fiducial absolute magnitude (a single constant for a particular
method), and $\Delta_i$ is a shift so that the peak absolute
magnitude for the SN is given by $(M_0+\Delta_i)$.  For the
purposes of this paper, we have ignored $K$-corrections and
extinction in \ceqn(mu-model)\ (one can consider them to have
been accounted for in the $m_i$ estimates and their uncertainties).  
These corrections could potentially be important sources of
systematic error; the observing teams have gone to some
lengths to constrain the sizes of such errors.  Here we concentrate
on the possibility that systematic errors are introduced by the 
lightcurve fitting algorithms entering the analyses via
the shifts $\Delta_i$.  We will seek information about such errors by
comparing the shifts {\em across} methods, rather than through
analysis of the internal consistency of a particular method.

The various fitting methods differ in how $m_i$ is interpolated from
the observed (incompletely sampled) lightcurve, in the choice for
$M_0$, and in how the shifts $\Delta_i$ are determined from the
(multicolor) lightcurve shapes.  The MLCS method provides $\Delta_i$
directly from fitting to a family of templates parameterized by
$\Delta_i$.  For the TF method, $\Delta_i = \beta_{\rm TF}(\Delta
m_{15} - 1.1)$, with $\beta_{\rm TF}$ a constant determined by fits.
For both the MLCS and TF methods, $M_0$ is inferred through the use of
SNe Ia that have Cepheid distances, and the various parameters
specifying the shift as a function of the lightcurve shape are set by
analyses of low redshift SNe.  For the SF method, $\Delta_i =
\alpha(s-1)$, where $s$ is the above-mentioned stretch factor and
$\alpha$ is a constant estimated jointly with cosmological parameters
in fits to the entire survey.  $M_0$ is simply set at an arbitrary
value; accordingly, no attempt is made to infer $H_0$ using SNe
analyzed with the SF method. In principle, each of the quantities on
the right hand side of \ceqn(mu-model)\ has uncertainty associated with
it, and the resulting errors in the estimates for these quantities are
correlated.  But only the combination given by \ceqn(mu-model)\ appears
in cosmological fits, so the lightcurve fitting results can be
summarized by the best-fit absolute magnitude $\hat\mu_i$ and the total
$\mu_i$ uncertainty $\sigma_i$ for each SN.  These quantities, and the
shifts $\Delta_i$, are the focus of our study.

Figure 2 shows histograms of the shifts deduced from the MLCS (R98),
TF (R98) and SF (P98) methods for the observed SNe Ia.  Since the choice
of $M_0$ can vary from method to method, we do not expect the histograms
to be aligned.  However, differences in histogram shape would indicate
that the various methods are correcting SNe in different and possibly
inconsistent ways.  While the three methods claim to reduce the
dispersion in the magnitude-redshift relationship at low $z$, it is
clear from the figure that they produce rather different distributions
of shifts.  Although the SF method has been applied to a different set of
SNe Ia than the other methods, this alone cannot explain the obvious
differences between the shapes of the histograms (we note that 14 SNe
are common to all three methods).  Most striking is that
the distribution is extremely narrow for the SF method, indicating that,
by this measure, the P98 SNe Ia sample consists almost entirely of
standard candles, or that for this sample of SNe Ia, the adopted
brightness-decline rate relation is not valid. This suggests to us that
these methods may be sensitive to different aspects of the SN Ia
phenomenon.  A consequence of this is that if the properties of SNe Ia
change with redshift, the relationships between the $\mu$ estimates
produced by the three methods could be $z$-dependent.  A search
for such a dependence could thus provide information about redshift
dependence of SNe Ia properties. In the following section we use
exploratory methods to search for evidence of this and other kinds of
dependences.

\section{Sleuthing}
\label{sec:fit}

Our approach in this section is driven by our belief that it is not
sufficient to settle for the consistency of the final cosmological
inferences of the MLCS, TF and SF analyses.  We should expect
consistency between them (statistically) on a supernova-by-supernova
basis where such a comparison is possible.

Since R98 use two different methods to compute distance moduli for
their sample of 37 SNe Ia, we can compare the results and search for
systematic differences between them.  Both the TF and MLCS techniques
are calibrated using the same set of nearby SNe Ia in the Hubble
flow and there is only one set of observational data for each SN Ia;
consequently, the uncertainties for the two methods are highly
correlated.  Another comparison set is the group of 14 supernovae from
the Cal\'an Tololo survey that are included in both R98 and P98.  Since
all fitting methods make use of the same published light curves for
this sample of 14 events, the inferred quantities for this sample will
also be highly correlated.

\subsection{Pointwise Consistency}

In Figure 3, we compare the distance moduli measured with the different
techniques on common samples of SNe Ia.  In Figure 3a, we show
$\mu_\MLCS$ vs $\mu_\TF$ for the 10 high redshift SNe Ia analyzed in
R98, and in Figure 3b, we show $\mu_\MLCS$ vs $\mu_\SF$ for the 14
common Cal\'an-Tololo SNe Ia analyzed by both SF and MLCS methods.  The
error bars for $\mu$ are derived from the uncertainties in the
individual distance moduli for each supernova (R98,P98), except that we
have removed the contribution associated with the intrinsic
dispersion of the SNe Ia sample, estimated to be $\sigma_{\rm int} = 0.10$
at low $z$, $\sigma_{\rm int} = 0.15$ at high $z$, 
\footnote{The error due
to intrinsic dispersion in the SNe Ia sample is estimated to be
somewhat larger in P98; $\sigma_{int} = 0.17$.  For the purposes of
comparison we remove the smaller estimated value of the correlated
error.} 
and we have removed the contribution associated with the
peculiar velocity of the SNe Ia ($\sigma_v = 300$ in P98, $\sigma_v =
200$ in R98).  Both the errors in the distance modulus from intrinsic
dispersion in the sample and from the peculiar velocity of the
SNe are completely correlated among the different methods.  We
have not removed the correlations due to $K$-corrections, photometry
and extinction (e.g., Schmidt et al.\ 1998), because there is
insufficient published information for us to do so properly;
consequently, we have overestimated the uncorrelated portion of the
distance modulus error somewhat.

{}From Figure 3 it is clear that the estimates for the distance moduli
from the different methods are strongly correlated, as they should be.
However, there is more dispersion  in these plots than we would expect
based on the quoted errors.  A fit of a straight line of slope 1 gives
a $\chi^2/\nu$ (with $\nu$ the number of degrees of freedom) 
of 22.8/9 for Figure 3a and 21.2/13 for Figure 3b
indicating that there are errors associated with the analysis methods
that have not been accounted for.

We can pursue this type of comparison further with the R98 data where
all of the SNe Ia have been fully analyzed with two independent
methods.  In Figure 4 we compare the MLCS and TF estimates of various
quantities that are used in inferring the distance moduli of the SNe Ia
events.  For the 37 SNe 1a analyzed in R98, Figure 4$a$ shows the host
galaxy extinction, $A$, 4$b$ shows the correction to the absolute
magnitude, $\Delta$, and 4$c$ illustrates the peak apparent magnitude,
$m$, calculated with the MLCS and TF analysis methods.  (The individual
errors for the extinction and $\Delta$ are not published but can be
crudely estimated to be of order 0.1 magnitudes.)  Again, there is more
dispersion evident in these plots than might be expected from the
quoted or estimated errors except for the correlation plot of
$m_\MLCS$ versus $m_\TF$.  The peak apparent magnitudes inferred via
the two methods, which are the quantities most directly related to the
raw data,  are in excellent agreement.

\subsection{Redshift and Luminosity Dependence}


In Figure 5, we plot the difference $\Delta\mu \equiv \mu_\MLCS - \mu_\TF$
between the distance moduli determined
from MLCS and TF respectively, as a function of $z$.
The error bars for $\Delta\mu$ are derived from
the uncertainties in the individual distance moduli except that,
as described above, we have removed the contribution associated with
the intrinsic dispersion of the SNe Ia sample. 
Formally, we use 
$\sigma^2_{\Delta\mu}=\sigma^2_\MLCS+\sigma^2_\TF-2\sigma_{\rm corr}^2$
to calculate the error bars shown in Figure 5. 
Although the data are somewhat scattered at both high and low $z$,
Figure 5 shows that the MLCS and TF methods agree rather well at
low $z$, apart from significant dispersion ($\sigma\sim 0.2$
mag) but there are hints of disagreement at large $z$, where
the dispersion, at least, appears larger, and the mean may also be shifted.

While it is possible that the appearance of Figure 5 at large $z$
merely reflects small number statistics, Figure 6 suggests that 
the incompatibility between TF and MLCS could be systematic. In
Figure 6, we plot $\Delta\mu$ versus $M_{B}^{\rm AV}$, an estimated
absolute magnitude, defined by

\begin{equation}
M_B^{\rm AV} = (M_B^\MLCS + M_B^\TF)/2 \label{eq:M}
\end{equation}
where 
\begin{eqnarray}
M_B^\MLCS & = & m_B^\MLCS - \mu_\MLCS - A_B^\MLCS \nonumber \\
M_B^\TF & = & m_B^\TF - \mu_\TF - A_B^\TF
\end{eqnarray}
and $A_B^{\rm MLCS(TF)}$ is an estimate of the  extinction due to the host
galaxy in
the MLCS (TF) correction scheme.  For $z>0.15$, R98
provided all of the information necessary to calculate
$M_B^\MLCS$, $M_B^\TF$ and hence $M_B^{\rm AV}$, but at low $z$,
$M_B$ was only given for a subset of the SNe Ia in Hamuy et al.\ (1996a).~\footnote{The
zero point reference for $M_B$ may be somewhat different for the
high redshift and low redshift data which may account for the
fact that in Figure 6, the low redshift supernovae seem to be, on average,
less luminous by about 0.5 magnitudes.  Our conclusions are robust 
against a shift in the zero point of the magnitude scale for the low
redshift supernovae.}

According to Figure 6, the difference between $\mu_\MLCS$ and
$\mu_\TF$ appears to be correlated with the estimated intrinsic
brightness, $M_B^{\rm AV}$, at high $z$, but not at low $z$. (Recall that
the error bars on $\Delta\mu$ are overestimates, as explained above.)
A similar correlation is evident if $\Delta\mu$ is plotted against
$\Delta_\MLCS(\Delta_\TF)$, the difference in maximum absolute
magnitudes for the observed SNe Ia and the fiducial SNe Ia according to
the MLCS(TF) method. (R98 tabulates $\Delta_\MLCS$ and $\Delta_\TF$
for all SNe Ia in their sample.) Figure~6 suggests that, at high
$z$, one of the
analysis schemes, MLCS or TF, either under-corrects or over-corrects
for the luminosity variations in the SNe Ia sample. Since no such
systematic trend is evident at low $z$, we cannot know which method, if
either, yields the more accurate value of distance modulus.
It is relatively uncontroversial to say that the two methods are not
identical, either at low or high $z$, and hence must probe SNe Ia
physics in slightly different, and as yet ill-understood ways
(H\"oflich and Kholkov 1996, H\"oflich, Wheeler and Thielemann 1998).
Thus, the indications of $z$-dependence implied by Figures 5 and 6, while
still based on relatively few events, are not especially surprising.

A worrisome feature of Figures 5 and 6 is that imperfect corrections
for luminosity variations can alter the conclusions about $\om$ and
$\ol$ that we draw from these data. To illustrate this point, we
computed $1\sigma$ confidence contours in ($\Omega_\Lambda,\Omega_M$)
space with separate fits to intrinsically bright and intrinsically dim
SNe Ia; the results are shown in Figure 7.~\footnote{In preparing
Figure 7, we have included a contribution to the uncertainty arising
from dispersion in galaxy redshift using the technique described in
R98.} The separation into `bright' and `dim' was somewhat arbitrary,
and we have verified that making different choices does not affect the
overall conclusion.~\footnote{ For the plots shown, we have chosen
$M_B^{AV}< -19.45$ as intrinsically bright and  $M_B^{AV}> -19.45$ as
intrinsically dim for MLCS and TF data.  For the SF data, we separated
intrinsically bright from intrinsically dim using $\alpha(s-1) >0 $ and
$\alpha(s-1) <0 $, respectively. Note that $\alpha(s-1)$ can be
calculated from the information in Tables 1 and 2 of P98.}
The $1\sigma$ confidence level contours for the combined data (all
$M_B^{AV}$) are also shown as dashed curves.  Figure 7 indicates a
systematic difference between the cosmology favored by intrinsically
bright versus intrinsically dim SNe Ia when the MLCS method is used;
the effect is much less pronounced for TF and seems to be of the
opposite sign for the SF method. The trend may be understood if the
MLCS (SF) method tends to overestimate (underestimate) the luminosities
of intrinsically bright SNe Ia at high redshift. Such a trend for the
MLCS data  is also consistent with Figure 6.

The set of plots in Figures 2 through 7 indicate that the analysis
methods disagree in their inferences of $\mu, A$, and $\Delta$ at a
level that is not covered by the quoted errors.  We can only
speculate on the sources of the discrepancies.  However until these
methods are understood more systematically, it will be difficult to
avoid assigning additional systematic errors to the measured distance
moduli with sizes that reflect the systematic differences between the
methods, and this will weaken the statistical
significance of the results substantially.

\subsection{Validity of Phillips Relations}

So far we have been investigating the light curve fitting methods as
possible sources of systematic error.  Potentially, there are other 
effects that can mimic cosmology that are extremely difficult to
constrain with the present data.  The most pernicious, discussed at
some length by the observers themselves, is evolution of the SN Ia
population.  It is extremely difficult to put reliable quantitative
limits on evolution, and it cannot be excluded conclusively using the
currently available spectral and color information.  Furthermore, there
is already some evidence  in the current data that the high redshift
sample does not have the same properties as the low redshift sample.

The strongest evidence that the lightcurve corrections improve our
knowledge of the SN absolute magnitude would be a demonstration that
they reduce the dispersion
of the SN distance moduli about the best-fit cosmology.  
(For example, Riess, Press \& Kirshner 1996 showed that MLCS reduces
the dispersion about Hubble's Law for low $z$ SNe Ia.) To test this,
we have
compared the dispersion between the data and the predictions of the
best-fit cosmology with and without the corrections, $\Delta_i$, inferred
from the light curve fitting.  \footnote{We redetermine the best fit
cosmology when we remove the corrections.} 
We adopt the quantity
\begin{equation}
D^2 = {1\over N} \sum_i [\hat\mu_i - f(z_i;H_0,\Omega_m,\Omega_\lambda)]^2 
\end{equation}
as a measure of the dispersion,
where $\hat\mu_i$ is the estimated distance modulus for SN $i$, $z_i$
is its redshift, the function $f(z)$ is defined by
\ceqn(eq:mu), and $N$ is the number of SNe Ia in the sample.  We
compute $D$ separately for high and low $z$; the results are
given in Table~1.

For both MLCS and TF, which are calibrated on low $z$ SNe Ia,
we see that the dispersion of the low redshift data is reduced
substantially by incorporating the corrections derived from the relation
between light curve shape and luminosity at maximum brightness.  At
high redshift, no such improvement is seen. The dispersion of the
high z data about the best fit cosmology is virtually unchanged by the
incorporation of either the MLCS or TF corrections.

For SF, there is little evidence from Table I that the corrections 
reduce the dispersion in the data at all.  Recall that in the SF
parameterization, the relation between light curve width and luminosity
corrections is parameterized by $\Delta_{SF}= \alpha(s-1)$ where
$\alpha$ is inferred from a global fit to the data at all redshift.  As
was shown in Figure 2, the corrections $\Delta_{SF}$ are quite small so
it is not surprising that they do little to reduce the
dispersion in the data.  As stated before, the SF method finds little,
if any, correlation between light curve width and absolute luminosity
when averaged over all redshift.  What is startling is that the low
redshift sample used in P98 is almost identical to the ``peak
subsample'' of Hamuy et al. (Hamuy et al.\ 1996a).  As detailed in that
reference, that low redshift sample does show a significant correlation
between light curve width and peak luminosity.  If a strong correlation
is present in the low redshift sample and only a very weak correlation
is evident in the full sample, one is led to suspect that the correlation
is not present in the high
redshift sample; the large number of
high-redshift SNe leverage the joint fit.

\section{Accounting For Possible Evolution}
\label{sec:stat}

Both R98 and P98 assume implicitly that the same light curve fitting methods
may be applied at all redshifts sampled.  This assumption is
only valid if the light curve shape is correlated with peak luminosity
in the same way at both high and low redshift.  Given the evidence we
have presented that this may not be true, which indicates, at least
circumstantially, that the SNe Ia population evolves, we feel
it is necessary to explore whether the data published so far actually are able
to actually distinguish the effects of evolution from those of
cosmology.

Such effects fall under the rubric of ``systematic errors''---because
they are not ``random,'' their effects on one's final inferences are
difficult to account for in the conventional frequentist approach to
statistical inference.  However, both teams have adopted the Bayesian
approach for their final analyses (though not for all intermediate
stages of their analyses).  As noted by Jeffreys (1961), the Bayesian
approach is particularly apt for studying the effects of systematic
error because of its broader notion of uncertainty.  A Bayesian
probability density describes how probability is distributed among
the possible values of a parameter, rather than how values of the
parameter are distributed among some hypothetical population.  This
permits statistical calculations with quantities that are not ``random''
in the frequentist sense.  In particular, as Jeffreys noted, 
systematic errors are treatable simply by introducing parameterized
models for the errors and marginalizing (integrating over) the
extra parameters to obtain one's final inferences.  

This procedure, when followed blindly, has the potential to weaken
one's conclusions unjustifiably.  For example, one could simply
introduce a systematic dependence that is identical to the physical
dependence one is studying, but with a duplicated set of parameters.
This duplication would prevent useful constraints from being placed on
the parameters, since any measured effect could be ``blamed'' on the
duplicated systematic dependence.  Thus Jeffreys emphasized the need to
compare models with and without systematic error terms using the ratio
of the model probabilities, the {\em odds} favoring one model over
another.  The odds can be written as the product of the prior odds
(expressing information from other data, or possibly a subjective
comparison of the models) and a {\em Bayes factor} determined entirely
by the data, the models, and the sizes of the model parameter spaces.
If we know or strongly believe a systematic effect to be present
without consideration of the new data before us, then obviously the
systematic error model should be used; the prior odds would lead us to
this conclusion even if the Bayes factor is indecisive.  If we have no
strong prior evidence for a systematic error, one takes the prior odds
to be unity and relies on the data alone for determining if the effect
is present, taking the Bayes factor to be the odds.  An appealing
aspect of Bayesian model comparison is that the Bayes factor implements
an automatic ``Ockham's razor'' that penalizes models for the sizes of
their parameter spaces.  Thus model complexity is accounted for by the
Bayes factor.  Except in unusual cases, needlessly increasing a model's
complexity by simply duplicating terms prevents the Bayes factor
from favoring the more complicated model.  We
provide a brief review of Bayes factors in Appendix~A; standard
references reviewing their use are Kass and Raftery (1995) and
Wasserman (1997).

\subsection{Systematic Error in $H_0$}

To illustrate this approach, we show how it can be used to
quantitatively account for systematic error introduced by the uncertain
Cepheid distances used to infer $M_0$ in the MLCS and TF methods.  (The
SF analysis used a ``Hubble-constant-free'' parameterization and thus
could avoid explicit treatment of $M_0$ and the Hubble constant.)  We
write the true value of $M_0$ as $(\hat M_0+\delta)$, where $\hat M_0$
is the estimate used for calculating $\hat\mu_i$, and the new term,
$\delta$, represents the constant (but unknown) error introduced by
using Cepheid data to calculate $\hat M_0$.  We describe the likelihood
function for analyzing the SNe Ia data in some detail in Appendix B.
The final (approximate) likelihood is equivalent to what one would
find from modelling the tabulated $\hat\mu_i$ estimates according to,
\bea
\hat\mu_i &=& f(z_i) + \delta + n_i\\
  &=& g(z_i) - \eta + \delta + n_i
\eqn(mu-delta)
\eea
where $f(z_i)$ is the cosmological distance modulus relation
defined in \ceqn(eq:mu), and $n_i$ is a random error term
whose probability distribution is a Gaussian with zero mean
and standard deviation $\sigma_i$.  In the second line, we
have separated out the $H_0$ dependence of $f(z_i)$ into
\be
\eta\equiv 5\log\left(h \over c_2 \right) - 25,
\eqn(eta-def)
\ee
where $H_0 = h\times 100\hbox{km}\,\hbox{s}^{-1}\,\hbox{Mpc}^{-1}$,
and $c_2$ is the speed of light in units of $10^2\hbox{km}\,\hbox{s}^{-1}$;
$g(z_i)$ contains the remaining $\OM$- and $\OL$-dependent part of $f(z_i)$.

It is clear from \ceqn(mu-delta)\ that $\eta$ (and thus $H_0$)
is degenerate with $\delta$; we cannot hope to learn about one
without independent knowledge of the other.  But $\delta$ is
constrained by our knowledge of the uncertainty of the Cepheid
distance scale.  In particular, R98 summarize the uncertainties
as introducing an error with a standard deviation of $d=0.21$ magnitudes
(corresponding to $\approx 10$\% uncertainty in $H_0$).
We account for this by introducing a prior distribution for $\delta$
that is a zero-mean Gaussian with standard deviation $d$.

Our model now has four parameters, $\delta$, $h$, $\OM$, and $\OL$.
The likelihood function for the data is the product of $N$ Gaussian
distributions specified by \ceqn(mu-delta)\ and is proportional to the
exponential of a familiar $\chi^2$ statistic.  The full joint posterior
distribution is the product of this and priors for the parameters,
including the informative prior for $\delta$.\footnote{The priors for
$\OM$, and $\OL$ we take to be flat over the region shown in our
plots, excluding the ``No Big Bang'' region; the prior for $h$
we take to be flat in the logarithm.}  We can summarize our
inferences for the cosmological parameters by integrating over
$\delta$; this can be done analytically and is described in
Appendix~B.  If we want to focus on the conclusions for $h$, we
numerically integrate over $\OM$ and $\OL$.  The result, for the MLCS
data, is the marginal distribution for $h$ shown as the rightmost solid
curve in Figure~8.  The best-fit value is $h=0.645$, and a 68.3\%
credible region has a half-width $\sigma_h = 0.063$.  This is
approximately equal to the ``total uncertainty'' on $H_0$ estimated by
R98 using standard ``rules of thumb'' for accounting for systematic
error; we have shown how this estimate could be justified by a formal
calculation.  For the TF data, the marginal posterior is plotted as the
leftmost solid curve in Figure~8, and $h= 0.627\pm 0.062$.

Of greater current interest are the implications for the density
parameters.  The marginal distribution for $\OM$ and $\OL$ is
found by integrating out $\delta$ and $h$.  This can be done
analytically (see Appendix~B).  Contours of the resulting
distributions, found using both the MLCS and TF data,
appear in Figure~9.  They are identical to
contours found using a model without $\delta$, and essentially
reproduce the results reported in R98 (minor differences result
from our omission of the ``snapshot'' SNe).

In this case, we know that $M_0$ has been estimated using the Cepheid
data, and that this estimate has systematic error.  Formally, the prior
odds favoring the model with $\delta$ over one with $\delta=0$ is thus
infinite.  The Bayes factor comparing these models is exactly equal to
one (this is because the SNe Ia data can tell us nothing about $\delta$; see
Appendix~A for discussion of this property of Bayes factors), so the
posterior odds is equal to the prior odds.  Since we know these errors
to be present, we take this $\delta$ model to be our ``default'' model
when calculating subsequent Bayes factors in this section.

We conclude our discussion of this model by summarizing the evidence in
the data for a nonzero cosmological constant, presuming the $\delta$
model to be true.  In R98 and P98, the marginal posterior probability
that $\OL>0$ was presented as such a summary; this probability was
found to equal 99.6\% (2.9$\sigma$), 99.99\% (3.9$\sigma$), and 99.8\%
(3.1$\sigma$) in the MLCS, TF, and SF analyses, respectively,
apparently indicating strong evidence that $\OL$ is nonzero.  But this
quantity is not a correct measure of the strength of the evidence that
$\OL \ne 0$.  This probability would equal unity if negative values of
$\OL$ were considered unreasonable a priori, 
yet presumably even in this case one
would not consider the data to demand a nonzero cosmological constant
with absolute certainty.  The correct quantity to calculate is the odds
in favor of a model with $\OL \ne 0$ versus a model with $\OL=0$.
Considering such models to be equally probable a priori, this is given
by the Bayes factor comparing these models.  
\footnote{These Bayes factor calculations can also be viewed as providing the
posterior probability that $\OL=0$ by putting a prior probability of
0.5 on the $\OL=0$ line; in the calculations reported in R98 and P98,
this line has zero prior probability (only finite intervals in
$(\OM,\OL)$ have nonzero prior probability in their analyses).}
We find $B=5.4$ using the MLCS data and $B=6.8$ using the SF data, each
indicating positive but not strong evidence for a nonzero cosmological
constant (presuming there is no evolution).  The TF data give $B=86$,
indicating strong evidence for a nonzero $\OL$ (again, presuming no
evolution).  Without clear criteria identifying one method as superior
to the others, the data are equivocal about a nonzero cosmological
constant, even without accounting for the effects of possible
evolution.

Similarly, R98 report the number of standard deviations that the
$\OM=1$, $\OL=0$ point is away from the best-fit $(\OM,\OL)$ as a
measure of the evidence against the hypothesis that matter provides the
closure density; they state this hypothesis is ruled out at the
$7\sigma$ and $9\sigma$ levels using the MLCS and TF methods,
respectively.  Again, a proper assessment of the hypothesis that
$\OM=1$ and $\OL=0$ requires that one give this hypothesis a finite,
nonzero prior probability.  For the MLCS method, the Bayes factor
favoring a model with any $(\OM,\OL)$ over one with $\OM=1$ and $\OL=0$
is $B=2.3\times10^4$, so that with equal prior odds the probability for
the latter model is $p=1/(1+B)\approx 5\times10^{-5}$ ($\approx
4.5\sigma$).  For the TF method, we find $B=2.1\times10^7$, so that
$p\approx 5\times10^{-8}$ ($\approx 5.8\sigma$).  These are small
probabilities and indicate very strong evidence against the simpler
model, but they are much larger than the probabilities associated with
$7\sigma$ and $9\sigma$ significances ($\approx 2\times10^{-11}$ and
$3\times10^{-18}$, respectively, for two degrees of freedom).  The
incorrect summary statistics used in the previous analyses have
exaggerated the evidence for a nonzero cosmological constant,
irrespective of whether or not one considers the effects of evolution.

\subsection{Models With Evolution}

Without a detailed physical idea of the cause of evolution, we cannot
explore truly realistic models.  Instead we consider two illustrative
examples.  We first consider a model (Model I) that generalizes the
$\delta$ model just described by adding an additional offset,
$\eps$, for the high redshift SNe; we
apply this model only to data from R98.  As a model of physical
evolution, this is certainly too simple, but it is illustrative since
for the R98 sample, the observed high redshift SNe Ia are all in a
fairly narrow band in redshift near $z\sim 0.5$.  Essentially this
model merely permits differences in luminosities between $z
\lesssim 0.1$ and $z \sim 0.5$ as a consequence of evolution.  On a purely
phenomenological level, the model might be considered more realistic
because the low and high redshift SNe are not treated equally in the
R98 analysis: the MLCS and TF relations are calibrated using only low
redshift SNe.  Thus this model can be understood as allowing for a
systematic offset when extrapolating the methods beyond the training
set.  For this model, we use the same prior for $\delta$ as in our
default model (zero-mean Gaussian with standard deviation $d=0.21$ mag).
The prior for $\eps$ we also take to be a zero-mean
Gaussian but with a different width $e$.  
The prior width, $e$, can be viewed as 
a description of the scale of errors we
might expect from extrapolating low redshift properties to high
redshift.  

Physically, we might expect evolution to lead to continuous variation
of SN Ia properties with redshift.  Also, the P98 analysis uses low and
high redshift SNe Ia together in calibrating the luminosity/decline
rate relation, so there is no clear separation of their data into low
and high $z$ subsamples.  Thus, Model~I is inappropriate for
phenomenological modeling of systematic effects from lightcurve fitting
of P98 data.  Consequently, we consider a second model (Model~II) which
assumes that the intrinsic luminosities of SNe Ia scale like a power of
$1 + z$ as a result of evolution.  This second model corresponds to
replacing \ceqn(mu-delta)\ with
\be
\hat\mu_i 
  = g(z_i) - \eta + \delta + \beta \ln (1+z) + n_i,
\eqn(mu-db)
\ee
where $\delta$ again represents Cepheid uncertainty in $M_0$ (relevant
only when we apply this model to MLCS or TF data), and $\beta$
parameterizes the evolution.  We use a Gaussian prior for $\beta$ with
zero mean and standard deviation $b$.   

For both models, we explore the dependence of the results on the prior
width ($e$ and $b$) to see how external constraints on evolution
(presently unknown) could affect the analysis.  We examine values that
allow evolutionary changes of up to a few tenths of a magnitude for
sources with $z\sim 1$.  These characteristic magnitude shifts are
comparable to the intrinsic dispersion seen in low redshift SNe Ia
(Schmidt et al.\ 1998), which may be taken as a rough indication of the
range of variation of peak magnitude with physical conditions in the
explosions.  Some current theoretical studies of possible sources of
$z$-dependent variations in SNe Ia luminosities also find magnitude
changes of this size to be reasonable (see, e.g., H\"oflich, Wheeler,
and Thielemann 1998; Dom\'inguez et al.\ 1999).

The new parameters in both models appear linearly in the model
equations and can thus be marginalized analytically.  Appendix~B
describes the calculations.  Reality could be and probably is far more
complicated than either model, but the sparsity of the present data do
not justify consideration of more sophisticated models.

\subsubsection{Model I}

Figure~10$a$ shows contours of the marginal density for $\OM$ and $\OL$
using Model~I to analyze the MLCS data and taking the prior width
for $\eps$ to be $e=0.1$ mag. Figure~10$b$ shows similar
results using the TF data.  It is clear from these figures that the
presence of a redshift-dependent shift of order $e$ greatly weakens our
ability to constrain $\OM$ and $\OL$ from the SNe Ia data.  The Bayes
factor for this model over the default $\delta$ model is $1.1$ for 
both MLCS and TF; the data alone are indecisive about whether such a
redshift-dependent error is present.
\footnote{The $\chi^2$ values for the default fits are already
acceptable, so one might worry that the more complicated models are
``overfitting.'' But the maximum likelihoods for models I and II are
only slightly greater than those found with the default model.  Bayes
factors account for overfitting and it is not playing a role here.  The
operation of Bayes factors is discussed further in Appendix A.}
Presuming it is present, the Bayes factor favoring nonzero $\OL$ over
$\OL=0$ is reduced significantly from what is found using
the default model; it is $1.1$ for MLCS and $3.8$ for TF.

These results are sensitive to our knowledge of the evolution.
Figures~10$c$ and 10$d$ show the MLCS and TF results again, but
this time with $e=0.2$ mag; the credible regions have grown
even larger in size.  Now the Bayes
factor for Model~I over the default $\delta$ model is $1.3$ for MLCS
and 1.2 for TF; the data remain indecisive about the presence of
an evolutionary offset.  Presuming it is present, the
Bayes factor favoring nonzero $\OL$ over $\OL=0$ is $0.7$ for MLCS 
(i.e., slightly favoring $\OL=0$) and $1.2$ for TF.  

As one might expect from these results, the most likely value of $\eps$
tends to be positive, making the more distant sources dimmer than the
nearer ones due to evolution rather than cosmology.  For example, in
Figures~10$c$ and 10$d$, when $\OM=1$ and $\OL=0$ (a point within the
95.4\% credible regions), $\eps=0.31\pm0.06$ for MLCS and
$\eps=0.17\pm0.06$ for TF.

The constraints placed on $H_0$ by the SNe Ia data arise mostly from
the low redshift objects, so one would not expect allowance for
evolution to drastically affect the $H_0$ inferences.  The analysis
bears this out.  The dashed curves in Figure~8 show the marginal
distributions for $h$ based on the MLCS and TF data using Model~I with
$e=0.2$; they differ little from the distributions found using the
reference model with no evolution.  Similar results are found using
Model~II.

\subsubsection{Model II}

Figure~11 shows results from analysis of the SF data using Model~II.
Figure~11$a$ shows contours of the marginal density for $\OM$ and $\OL$
for the $\beta=0$ case as a reference; these contours essentially
duplicate the results of Fit C in P98.  Figure~11$b$ shows similar
contours, but allowing for a nonzero $\beta$; the prior standard
deviation for $\beta$ was $b=0.25$.  Figure~11$c$ repeats the
analysis with $b=0.5$.  Again we find that the possibility
of evolution significantly weakens the constraints on the density
parameters, but if the amount of evolution can be bounded, useful
limits might result.  The Bayes factor for evolution vs.\ no
evolution is $1.0$ for $b=0.25$ and $1.1$ for $b=0.5$, so the
data alone are indecisive about the presence or absence of this type of
evolution.  We find similar results when using this model to analyze
the MLCS and TF data.

Figure~12 shows how these findings depend on the prior uncertainty
for $\beta$.  The solid curve shows how the Bayes factor favoring
a nonzero $\OL$ over $\OL=0$ depends on $b$; only for $b\lesssim0.1$
does the Bayes factor remain near the value of 6.8 found assuming
there is no evolution.  The dashed curve shows the Bayes factor
for Model~II versus the default model with no evolution; for
no value of $b$ in the range of the plot can the data clearly
distinguish evolution from cosmology.  This emphasizes the need
to use information independent of the $\mu$-$z$ relation to
constrain evolution.

\subsubsection{Flat Cosmologies}

So far we have assessed the evidence for nonzero $\OL$ by
comparing models with $\OL=0$ to models with arbitrary
$\OL$, as was done in the R98 and P98 analyses.  However,
many cosmologists would consider flat models, with
$\OL = 1 - \OM$, to be of special relevance (e.g., because
of inflationary arguments).  We have thus analyzed models
constrained in this way, using our default model and models
I and II.

Figure~13 shows the marginal posterior distributions for $\OM$ (and,
equivalently, for $\OL = 1 - \OM$) presuming a flat cosmology.  The
three panels show analyses of the distance moduli reported using the
MLCS and TF data with Model~I (top and middle,
respectively), and using
the SF data with Model~II (bottom).  The solid curves
show results based on the default model; the short-dashed curves show
results with a small amount of evolution allowed ($e = 0.1$ or
$b=0.25$), and the long-dashed curves show results with a larger amount
of evolution allowed ($e = 0.2$ or $b=0.5$).  As in the previous cases,
the Bayes factors comparing models with evolution to the default model
are all nearly equal to one.  Also, as was found before, accounting for
the possibility of evolution significantly weakens the evidence for
nonzero $\OL$.  However, if one restricts attention to flat models, the
evidence for nonzero $\OL$ is stronger than it is if one allows nonflat
cosmologies.  For the default model presuming no evolution, the Bayes
factors favoring nonzero $\OL$ over a flat model with $\OM=1$ are
$2.1\times10^4$ (MLCS), $2.5\times10^6$ (TF) and $5.0\times 10^3$ (SF),
much larger values than were found in the comparison using nonflat $\OL$
models discussed above.  But these values fall dramatically if one
allows for evolution.  For models with a small amount of evolution
allowed, they decrease to 20, 48, and 14, respectively, indicating
positive but not compelling evidence for nonzero $\OL$.  For models
with a larger amount of evolution allowed, they decrease further to
2.4, 2.5, and 2.3, indicating no significant evidence for nonzero
$\OL$.

\subsection{Simulations}

Figure~14 elucidates why introducing the possibility of evolution
weakens our ability to constrain $\OM$ and $\OL$ so greatly.  The thick
solid curve shows $g(z)$ for the best-fit cosmology from fits to the SF
data presuming no evolution ($\OM=0.75$, $\OL=1.34$).  The dotted curve
shows the same function for the flat $\OM=1$ ($\OL=0$) cosmology (which
lies within the 68.3\% credible region in Figure~11$c$), and the dashed
curve shows the evolutionary component $\beta\ln(1+z)$ for the best-fit
value of $\beta$ given this cosmology ($\beta = 0.83$).  The thin solid
curve shows the sum of the dotted and dashed curve.  Over the range of
redshift covered by the data ($z\lesssim 1$) and even beyond,
$(\OM,\OL,\beta) = (0.75, 1.34, 0)$ and $(\OM,\OL,\beta) = (1,0,0.83)$
are indistinguishable if one allows evolution of this type, unless one
can determine $\mu$ to significantly better than the $\sim20$\%
accuracy currently obtained at $z\lesssim 1$.  However, we note that
the models differ substantially at larger redshift (by about 0.8 mag),
which offers some hope of discerning evolution. We caution, though,
that the best fit values of $(\OM,\OL)$ are likely to be different for
data extending to $z\approx 2$, either with or without evolution, so
the comparison is not truly apt, and moreover $(\OM,\OL)=(0.75,1.34)$
might be considered implausible intrinsically by many cosmologists.  To
amplify this point, we also compare $(\OM,\OL,\beta)=(1,0, 0.83)$ with
another cosmology, $(\OM,\OL)=(0.3,0.7)$, within the 68.3\% credible
region of the no-evolution analysis.  As is shown by the bold,
long-dashed curve in Figure~14, $\mu$ accuracies better than 10\% out
to $z \approx 2$ would be needed to distinguish these cosmologies from one
another out.  We have systematically explored a wide
range of cosmologies and found similar results:  simple power law
evolution can make widely disparate cosmologies appear remarkably
similar.  Put another way, the differences between cosmologies with
various $\OM$ and $\OL$ are well mimicked by power-law evolution to
redshifts beyond those currently accessible in supernova surveys.  We
emphasize that we did not choose the form of the evolution to produce
this degeneracy; this is a standard phenomenological model for
evolution.  We have found similar behavior with another simple model
for evolution consisting of a power law in lookback time.

To assess how well evolution can mimic cosmology, we analyzed simulated
data consisting of $\mu$ values with added Gaussian noise at redshifts
that themselves had added Gaussian noise.  The simulations were
designed to roughly mimic possible future data like that reported in
P98 (this simplified the analysis since $H_0$ need not be accounted for
explicitly as it would have to be for data like those reported in R98).
The redshifts of the first 16 SNe in our simulated data sets were
chosen to be similar to those of the 16 low-redshift SNe in P98
($z\lesssim 0.1$); the redshifts for the remaining simulated data were
chosen randomly from a uniform distribution over some specified
interval.  We added redshift errors with a standard deviation of 0.002,
and $\mu$ errors with standard deviations equal to those reported by
P98 for the 16 low-$z$ SNe, and equal to 0.25 magnitudes for the
high-$z$ SNe (a typical value for $\mu$ values reported in P98).

Figure~15 shows typical results.  Here we simulated data from a flat,
$\OM=1$ ($\OL=0$) cosmology with evolution described by Model~II with
$\beta=0.5$.  Figure~15$a$ shows the results of an analysis assuming no
evolution, with 38 simulated SNe redshifts in the interval $[0.3,1]$
(54 total simulated points).  This corresponds to a sample size equal
to that used in the P98 analysis and extending over a similar range of
redshift.  The cross indicates the best-fit parameter values, the dot
indicates the true values, and the contours bound credible regions of
various sizes.  One would reject the true model as being improbable if
evolution is ignored.  Figure~15$b$ shows a similar plot, with the
number of high-$z$ SNe increased so the total sample size is now 200,
with the high-$z$ points now spread over $[0.3,1.5]$.  The contours
have shrunk considerably, converging around a point well away from the
truth.  In both figures, the best-fit point has an excellent
$\chi^2/\nu$ ($53.6/52$ for the small data set, $201/198$ for the large
one).  Evolution mimics cosmology so well that standard ``goodness of
fit'' reasoning can lead one to conclude, mistakenly, that pure
cosmology (with no evolution) is an adequate description of data of
this quality even when substantial evolution is present.

Figure~15$c$ shows the results
of an analysis of the larger data set using a model that
includes evolution; the marginal posterior (with $\beta$ 
marginalized) is shown.  The credible regions now contain
the true model, but they are large even for a data set four
times the size of the currently published surveys, and
extending to significantly higher redshift.  The Bayes factor
is of order unity, showing that data of this quality is
not capable of distinguishing between models with and
without evolution. This is further testimony to the approximate
degeneracy between cosmology and evolution, at least at
$z\lesssim 1.5$.

The extent to which evolution corrupts the results depends both
on the true cosmology and on the amount of evolution allowed.  
Independent constraints
on the amount of evolution could thus play an important role
in allowing useful contraints to be placed on the cosmology.
They would enter the analysis via the prior for $\beta$.
Comparison of Figures~11$a$ through 11$c$ shows how constraints on
the amount of evolution can affect one's final inferences.

\section{Conclusions}
\label{sec:conc}

Systematic uncertainty may enter the analysis of any data set as
a result of real physical effects that are not accounted
for explicitly.  
As an example,  the use of observations 
of distant galaxies
to measure the cosmological deceleration parameter had to 
confront the systematic
errors introduced by the
fact that not only are galaxies not standard candles, but
their luminosities also evolve with time (e.g., Tinsley 1968;
Weinberg 1972; Ostriker \& Tremaine 1975; Tinsley
1977; Sandage 1988; Yoshii \& Takahara 1988; Bruzual 1990; Peebles 1993).
A principal goal of this paper 
has been to present a study of the systematic error due to evolution
in attempts to determine $\OM$ and $\OL$ from
observations of SNe Ia. 
   
One focus of this paper has been to see if
there are indications that the SNe Ia population has evolved
from $z\sim 0.5-1$ to $z\ll 1$. We have presented two arguments
that this might be so. The first is that a comparison of
the peak luminosities estimated for individual SNe Ia by
two different methods, MLCS and TF, are not entirely
consistent with one another at high redshifts, $z\sim 0.5$.
We have asserted that
the two methods very likely sample slightly different
aspects of the SN Ia mechanism, and should not be expected
to agree completely. If evolution were entirely absent,
though, the differences between them should not depend on
redshift, contrary to the admittedly sketchy evidence of
the data. A second hint that SNe Ia evolve with redshift 
is that while the three luminosity estimators, SF as well as
MLCS and TF, reduce the dispersion of distance moduli
about best fit models at low redshift, they do not at
high redshift.  

These studies were intended to give us impetus to pursue
the more fundamental point of this paper, namely that
evolution {\it must} be considered possible, even if
there are no ``smoking guns'' that seem to require it.
Ideally, one should attempt to constrain the parameters of
an evolutionary model at the same time as determining
the parameters of the cosmological model. As we stated
at the outset, changes in peak SN Ia magnitude of order 0.1
magnitudes out to $z\sim 1$ would alter the ranges of
acceptable cosmological models substantially. 
The dispersion of SNe Ia peak magnitudes at low $z$ is approximately
0.3-0.5 mag (Schmidt et al. 1998), which might indicate a plausible
range of variation for diverse physical conditions. Using theoretical
models, H\"oflich, Wheeler \& Thielemann (1998) argue that a similar 
range of variation of peak luminosities could arise as a consequence
of changes in composition which might be due to evolution.

To get an idea of how allowing for the possibility of
evolution would affect one's
ability to constrain cosmological parameters, we considered
two different models. In one, we assumed that there is
a constant magnitude shift between low and high redshift. 
We also considered a model in which the
peak magnitudes of SNe Ia evolve continuously, with
$\delta m(z)=\beta\ln(1+z)$.
In applying these models, prior
assumptions about the amplitude of possible magnitude
changes of SNe Ia between low and high redshifts are 
needed to evalute the systematic error that might be
introduced.  At present, little is known, so our
calculations allow a range of possibilities.  To
do this, we assume Gaussian prior probability distributions for the
(unknown) parameters of the evolutionary models.  These priors
express a preference for no evolution, but have adjustable standard
deviations that encapsulate prior notions about how large possible evolutionary
effects might be. The results presented in P98 and R98 correspond to setting 
these standard deviations to zero; i.e., no
evolution at all. We adopt a more conservative viewpoint, and present
results for different choices of the ranges of magnitude evolution
that are allowed {\it a priori}. Significantly, when we permit peak magnitude
changes out to $z\sim 1$
comparable to (and even somewhat smaller than) the range observed
for low redshift SNe Ia (Schmidt et al. 1998), the implied systematic
uncertainty in $\Omega_M$
and $\Omega_\Lambda$ becomes so large that the data cannot constrain
these cosmological parameters usefully.
However, our ability to determine $H_0$ is virtually unaffected by evolution.

In order to assess the extent to which the data favor
models allowing evolution over ones without evolution, we computed
Bayes factors. The Bayes factor between classes of models
with and without luminosity evolution are equivalent
to the odds ratio between them if there is no {\it
a priori} reason to prefer one over the other. In
all cases, we found that the Bayes factors are
of order unity, which means that the data themselves
do not favor either model. If we accounted for
a prior prejudice that evolution does occur, the odds
would disfavor models in which the SNe Ia population
has the same properties at all redshifts. 

The two models we have considered illustrate how well evolution can
mimic cosmology. The less realistic model merely allowed a shift
in the magnitudes
of high $z$ SNe Ia relative to low $z$ SNe Ia by a fixed (but uncertain) amount,
$\delta m$. Since the SNe Ia in the R98 sample were predominantly at
$z\approx 0.3-0.5$, the cosmological magnitude shift (relative to
Hubble's law or any other fiducial cosmology) varies little over the
entire redshift range they span. Clearly, for the high $z$ SNe Ia in
this sample, one only knows that there is a total magnitude shift
between $z\lesssim 0.1$ and $z\sim 0.5$, not how much is due to
cosmology and how much to evolution.  The characteristic magnitude
shifts due to evolution needed for a cosmological model with
$(\OM,\OL)=(1,0)$ are $\approx 0.2-0.3$.  Ultimately, a model in
which there is simply a constant magnitude difference between low and
high $z$ SNe Ia should fail to model data spanning a lage range of
redshifts. 

More daunting is the success of models allowing a continuous
magnitude shift, $\delta m(z)=\beta\ln(1+z)$. While it
is unsurprising that such models would be approximately degenerate
with cosmology at low $z$, where the combined magnitude
shift, relative to Hubble flow, is
$[1.086(1-\Omega_M/2+\Omega_\Lambda)+ \beta]z$
(e.g., Weinberg 1972),
it is remarkable that a continuous magnitude shift with
this simple form cannot be discerned out to at least
$z\approx 1$. Our simulations show that even if there is
truly no evolution, so that reality corresponds to certain
values of $\Omega_M$ and $\Omega_\Lambda$ with $\beta=0$,
models with $\beta\neq 0$ and $(\Omega_M^{\rm eff},
\Omega_\Lambda^{\rm eff})\neq(\Omega_M,\Omega_\Lambda)$ yield
apparent magnitudes that are indistinguishable from the
truth within differences in distance modulus
$\approx 0.1$ mag.
Differences between select cosmological models may be larger
at higher $z$, but often remain within $\sim 0.1$ mag out to
$z\approx 2$.

We also use simulations to explore the converse situation where we
neglect evolution in the analysis of samples of evolving SNe Ia.
As an example, we simulated a set of 200 SNe Ia distance moduli,
including 184 high-$z$ SNe with redshifts uniformly
distributed over $0.3\leq z\leq 1.5$, in a cosmological model with
$(\Omega_M,\Omega_\Lambda,\beta)=(1.0, 0.0, 0.5)$.  We then analyzed
the data with evolution neglected entirely.  The result was that given
enough SNe Ia, the analysis would pick out a small range of ``allowed''
values of $(\Omega_M,\Omega_\Lambda)$, but centered around incorrect
values. The true cosmology was well outside the $3\sigma$ credible
region for these simulations, yet the (incorrect) best-fit model would
be judged excellent by a standard $\chi^2$ goodness-of-fit test.

What is needed to separate evolution from cosmology is both
detections of greater numbers of SNe Ia at high redshift
with detailed measurements of light curves and spectra, and,
equally important, a
better physical understanding of the SN Ia process. In particular, one
would like to be able to link the Phillips relations,
lightcurve risetimes and spectra, uniquely, to
internal conditions in the explosions themselves, to be able to
understand how they might evolve with redshift (see, e.g., 
von Hipple, Bothun, \& Schommer 1997; H\"oflich,
Wheeler, and Thielemann 1998; Dom\'inguez et al.\ 1999).  This would
allow construction of realistic, not phenomenological, models for
evolution, and one might hope to be able to constrain the parameters of
these models along with cosmological ones. The analogue in galactic
astronomy is the use of population synthesis models to study the
cosmological evolution of the luminosity function, which {\it might}
permit, given enough data, simultaneous fits for cosmological
parameters (Yoshii \& Takahara 1988, Bruzual 1990). Such
detailed physical modelling might lead to a detailed, quantitative
connection between the peak luminosities of SNe Ia and their spectra,
which would allow additional information to be useful quantitatively in
fitting for $\Omega_M$ and $\Omega_\Lambda$. R98 and P98 have argued,
using the spectral data, that there is no compelling evidence {\it for}
evolution, but that does not translate into a convincing argument {\it
against} evolution unless the salient features of the source spectra
can be connected unambiguously to peak luminosity.
In fact, since this paper was submitted,
Riess, Filippenko, Li \& Schmidt (1999) have claimed that the
rise times of low and high redshift SNe Ia are different even
though earlier studies found no comparably strong evidence for spectral
evolution.

In the end, what all cosmologists want to know is the probability that
the cosmological constant is nonzero.  The Bayes factor provides
straightforward mathematical machinery for doing this calculation,
whether or not evolution is included in the analysis.  When
the possibility of evolution
is not included in the analysis, and no prior assumptions are made
about the spatial geometry of the Universe, the Bayes factor for
$\Omega_\Lambda\neq 0$ compared to $\Omega_\Lambda=0$ is $B=5.4$ using
the MLCS method, $B=6.8$ using the SF method, and $B=86$ using the TF
method, which if one is not prejudiced either way, only favors nonzero
$\Omega_\Lambda$ equivocally. (There may be reasons to be prejudiced
one way or the other; see for example Turner 1999 for a theoretical
cosmologist's point of view.) When the possibility of evolution is 
accounted for in the analysis, the
values of the analogous Bayes factors depend on one's prior
assumptions, but rather conservatively $B\lesssim 1$.  Thus, if we do
not discriminate among open, closed and flat cosmological models, the
data alone do not choose between $\OL\neq 0$ and $\OL=0$ once the
possibility of evolution is taken into account. However, if the
Universe is presumed to be flat spatially, then the case for $\OL\neq
0$ is stronger. If evolution is presumed not to occur, we find Bayes
factors $B=2.1\times 10^4$ (MLCS), $2.5\times 10^6$ (TF) and $5.0\times
10^3$ (SF), decisive odds in favor of nonzero $\OL$. Weak evolution
($e=0.1$ in Model I or $b=0.25$ in Model II) lowers these values to
$B=20$ (MLCS), 4.8 (TF) and 14 (SF), which still favors $\OL\neq 0$
positively but not nearly as persuasively. If evolution is allowed
to be somewhat
more pronounced, but still at a plausible level ($e=0.2$ or $b=0.5$),
the Bayes factors fall to $B=2.4$ (MLCS), 2.5 (TF) and 2.3 (SF), which
is scant evidence for a non-vanishing cosmological constant. Once
again, the ability of the data to distinguish $\OL\neq 0$ from $\OL=0$
depends sensitively on prior assumptions about evolution of SNe Ia, and
underscores the importance of placing independent constraints on the
possible range of variation of their peak luminosities with redshift.

\acknowledgments

We have benefited from conversations and correspondence with Adam
Riess, Peter Garnavich and Bill Press.  We thank Saul Perlmutter
for providing a draft of P98 prior to publication.
We thank Margaret Geller, Saul Teukolsky, \'Eanna Flanagan, 
David Chernoff, Ed Salpeter, and Sidney Drell for comments on
earlier drafts of this paper.
PSD gratefully acknowledges support from the John Simon Guggenheim
Memorial Foundation.  This work was supported by NASA grants NAG5-3097,
NAG5-2762, NAG5-3427, and NSF grant PHY-9310764.

\appendix
\section{Bayes Factors}

In Bayesian inference, to form a judgement about an hypothesis
$H_i$, we calculate its probability, $p(H_i|D,I)$, conditional
on the data ($D$) and any other relevant information at hand ($I$).
The desired probability $p(H_i|D,I)$ is not usually assignable directly;
instead we must calculate it from other simpler probabilities using the
rules of probability theory.  Prominent among these is Bayes's
theorem, expressing this posterior (i.e., after consideration of the
data) probability in terms of a prior probability for $H_i$
and a likelihood for $H_i$,
\be
p(H_i|D,I) = p(H_i|I) \, {\like(H_i) \over p(D|I)},\eqn(BT-H)
\ee
where the likelihood for $H_i$, $\like(H_i)$, is a shorthand notation
for the sampling probability for $D$ presuming $H_i$ to be
true, $p(D|H_i,I)$.  The likelihood notation and terminology emphasizes
that it is the dependence of the sampling probability on $H_i$ (rather than
$D$) that is of interest for calculating posterior probabilities.
The term in the denominator is the prior predictive probability
for the data and plays the role of a normalization constant.  It
can be calculated according to
\be
p(D|I) = \sum_i p(H_i|I) \like(H_i).\eqn(pd-sum)
\ee
We see from this equation that the prior predictive probability
is the average likelihood for the hypotheses, with the prior
being the averaging weight.  It is also sometimes called the
marginal probability for the data.

For estimating the values of the parameters $\theta$ of some model, the
background information is the assumption that the parameterized model
under consideration is true; we denote this by $M$ (this may include
any other information we have about the parameters apart from that
provided by $D$; for example, previously obtained data).  The posterior
probability for any hypothesis about continuous parameters can be calculated
from the posterior probability density function (PDF), which we may
calculate with a continuous version of Bayes's theorem:
\be
p(\theta|D,M) = p(\theta|M) \, {\like(\theta) \over p(D|M)}.\eqn(BT)
\ee
Both the posterior and the prior are PDFs in this equation; we
continue to use the $p(\cdot)$ notation, letting the nature of the
argument dictate whether a probability or PDF is meant.  In this
case, the normalization constant is given by an integral:
\be
p(D|M) = \int d\theta\; p(\theta|M) \, \like(\theta).\eqn(pd-int)
\ee
The normalization constant is now the average likelihood for the
model parameters.

When comparing rival models, $M_i$, each with parameters $\theta_i$,
we return to the discrete version of Bayes's theorem in
\ceqn(BT), using $H_i = M_i$ for the hypotheses, and taking the
background information to be $I=M_1 + M_2 + \cdots$ (denoting the 
proposition, ``Model $M_1$ is true {\em or} model $M_2$ is
true {\em or} $\ldots$'').  The likelihood for model $M_i$
is $p(D|M_i,I)$; but since the joint proposition $(M_i,I)$ is
equivalent to the proposition $M_i$ by itself, we
have $\like(M_i) = p(D|M_i)$.  Thus the likelihood for a model
in a model comparison calculation is equal to the normalization
constant we would use when doing parameter estimation for that
model, given by an equation like \ceqn(pd-int).  In other
words, the likelihood for a model (as a whole) is the average
likelihood for its parameters.

It is convenient and common to report model probabilities via
odds, ratios of probabilities of models.  The (posterior) odds for $M_i$
over $M_j$ is
\bea
O_{ij}
  &\equiv& {p(M_i|D,I) \over p(M_j|D,I)}\nonumber\\
  &=& {p(M_i|I) \over p(M_j|I)}\times {p(D|M_i) \over p(D|M_j)}\nonumber\\
  &\equiv& {p(M_i|I) \over p(M_j|I)}\times B_{ij} 
\eqn(O-def)
\eea
where the first factor is the prior odds, and the
ratio of model likelihoods, $B_{ij}$, is called the
{\em Bayes factor}.  When the prior information does not
indicate a preference for one model over another, the
prior odds is unity and the odds is equal to the Bayes factor.
Kass and Raftery (1995) provide a comprehensive review of Bayes
factors, and Wasserman (1997) provides a survey of their use
and methods for calculating them.  When the prior odds
does not strongly favor one model over another, the Bayes
factor can be interpreted just as one would interpret
an odds in betting; Table 2 summarizes the recommended
interpretation of Kass and Raftery.

The Bayes factor is a ratio of prior predictive probabilities;
it compares how rival models predicted the observed
data.  Simple models with no or few parameters have their
predictive probability concentrated in a small part of
the sample space.  The additional parameters of complicated
models allow them to assign more probability to other
regions of the sample space, but since the predictive
probability must be normalized, this broader explanatory
power comes at the expense of reducing the probability
for data lying in the regions accessible to simpler models.
As a result, model comparison using Bayes factors
tends to favor simpler models unless the data are truly
difficult to account for with such models.  Bayes factors
thus implement a kind of automatic and objective ``Ockham's
razor'' (Jaynes 1979; Jefferys and Berger 1992).

This notion of simplicity is somewhat subtle, but in some
simple situations it accords well with our intuition
that models with more parameters are more complicated and should
only be preferred if they account for the data significantly
better than a simpler alternative.  Because Bayes factors
are ratios of {\em average} likelihoods, rather than the
{\em maximum} likelihoods that are used for model comparison
in frequentist statistics, they penalize models for the sizes of their
parameter spaces.  A simple, approximate calculation
of the average parameter likelihood given by \ceqn(pd-int)\ 
elucidates how this comes about.

First, we assume that the data are informative in the sense of
producing a likelihood function that is strongly localized compared to
the prior.  Suppose that the scale of variation of the prior is
$\Delta\theta$, and the scale of variation of the likelihood is
$\delta\theta \ll \Delta\theta$.  If the likelihood is maximized at
$\theta = \hat\theta$, then we find
\be
p(D|M) \approx p(\hat\theta|M) \int d\theta\;  \, \like(\theta).\eqn(pd-1)
\ee
Since the prior is normalized with respect to $\theta$,
$p(\hat\theta|M)$ will be roughly equal to $1/\Delta\theta$.  The
integral will be roughly equal to the product of the peak and width of
the likelihood, $\like(\hat\theta)\delta\theta$.  Thus,
\be
p(D|M) \approx \like(\hat\theta){\delta\theta\over\Delta\theta}.\eqn(pd-approx)
\ee
We find that the likelihood for a model is approximately given by the
maximum likelihood for its parameters, multiplied by a factor that is
always $\le 1$ that is a measure of how the size of the probable
part of the parameter space changes when we account for the data.  This
latter factor is colloquially known as the {\em Ockham factor}.  To see
why, consider the case of nested models:  $M_1$ and $M_2$ share
parameters $\theta$, but $M_2$ has additional parameters $\phi$.  In
such cases, it is not uncommon that the prior and posterior ranges for
$\theta$ are usually comparable for both models (this is not the case
in the present work, however).  Then the Bayes factor in favor of the
more complicated model is approximately given by
\be
B_{21} \approx {\like(\hat\theta,\hat\phi) \over \like(\hat\theta)}
  {\delta\phi\over \Delta\phi}.\eqn(B-nest)
\ee
Thus the data will favor $M_2$ only if the maximum likelihood ratio is
high enough to offset $\delta\phi\over \Delta\phi$, which will be $< 1$
if the data contain any information about $\phi$ (and cannot be $>1$ in
any case).  This is in contrast to the frequentist approach, where only
the ratio of maximum likelihoods is used.  This ratio cannot disfavor
$M_2$; one thus requires the likelihood ratio to exceed some critical
value before preferring $M_2$, on the grounds that one should prefer
the simpler model a priori.  Unfortunately, the critical value is set
in a purely subjective and ad hoc manner, and comparisons using
likelihood ratios can be inconsistent (in the formal statistical sense
of giving the incorrect answer when the amount of data becomes
infinite).  The Bayesian approach can (and often does) prefer the
simpler model even when both models are given equal prior
probabilities, and the critical likelihood ratio needed to just prefer
$M_2$ is determined by the likelihood functions and the size of the
parameter space searched.  The odds is known to be a consistent
statistic for choosing between models.

The approximations leading to the simple result of \ceqn(B-nest)\ are
not valid for the present work, so a simple ``Ockham's razor''
interpretation of our results is not possible.  Although the default
model is nested in the models that have $z$-dependent systematic
errors, it is clear from the
figures that the addition of the systematic error parameters
(corresponding to $\phi$ in the above analysis) greatly affects
inferences of the cosmological parameters (corresponding to $\theta$).
Thus the $\delta\theta$ factors (here associated with the cosmological
parameters) do not approximately cancel in the Bayes factor.  
Moreover, inferences for the $\theta$ and $\phi$ parameters
are highly correlated in the SNe Ia problem, so it is not 
possible to identify separate $\delta\theta$ and $\delta\phi$
factors separately quantifying the uncertainties in the
nested and additional parameters.  We do know that the maximum
likelihoods (e.g., minimum $\chi^2$ values) are comparable
for models with and without $z$-dependent systematic errors.
The more complicated models are not improving the best fit
substantially, but rather the additional parameter allows one
to make the fit nearly as good as the best fit throughout a large
region of the parameter space (because of the near-degeneracy
of evolution and cosmology).  It is this increase of the
acceptable volume of parameter space that accounts for the
Bayes factors slightly favoring the more complicated models here.

As is clear from \ceqn(pd-approx), the prior ranges for parameters
play an important role in Bayesian model comparison.  This is in
contrast to their role in parameter estimation, where in Bayes's
theorem the prior range factor appears in both the numerator (through
the prior) and the denominator (through the average likelihood) and
thus cancels, typically having a negligible effect on inferences
(though the range itself cancels, some effect can remain due to
truncation of the tails of the likelihood).  In particular, parameter
estimation is typically well-behaved even when one uses
improper (non-normalizable) priors, such as flat priors
with infinite ranges.  But model comparison fails when the priors for any
parameters not common to all models are improper, because the Ockham
factors associated with those parameters vanish.  This may at first
appear to be troubling (or at best a nuisance), but a similar dependence
on the prior range of parameters
is acknowledged to be necessary even in frequentist treatments of many
problems.  For example, consider detection of a
periodic signal in a noisy time series using a power spectrum estimator.  
This is a model comparison problem (comparing a model
without a signal to one with a periodic signal), and in fact the
spectral power is simply related to the likelihood for a periodic
(sinusoidal) signal.  In frequentist analyses, one
cannot simply use the number of standard deviations the spectral peak 
is above the null expectation to assess the
significance of a signal; one must also take into account the number of
statistically independent frequencies examined, which depends on the
frequency range searched and on the number and locations of frequencies
examined within that range.  Similar considerations arise in
searches for features in energy spectra, or searches for sources in
images---one must take into account the number and locations of points
searched in order to properly assess the significance of a detection.
The results of the corresponding Bayesian calculations similarly depend
on the ranges of parameters searched (but not on the number and
locations of the parameter values used).  Bayes's theorem indicates
that the sizes of parameter spaces (i.e., search ranges) must be taken
into account whenever we compare models; such considerations should not
be unique to the few applications where they have been recognized to be
important in conventional analyses.

\section{Statistical Methodology}

As in the analyses of R98 and P98, we adopt the Bayesian approach for
inferring the cosmological parameters $\OM$ and $\OL$, extending
their analyses to include parameterized systematic and evolutionary
components.  The additional parameters are dealt with by marginalizing
(as the R98 analysis did with $H_0$ and the P98 analysis did
with the SF fitting parameters).  Many of the needed marginalizations
can be done analytically; this Appendix describes these calculations.
Some remaining marginalizations (including calculation of Bayes
factors) were done numerically with various methods including
straightforward quadrature, adaptive quadrature, and Laplace's 
method; application of these methods to Bayesian integrals
is surveyed in Loredo (1999).

\subsection{Basic Framework}

Let $D_i$ denote the data associated with SN number $i$, and $D$
denote all the data associated with the $N$ SNe in a particular survey.
Let $\cpar$ denote the cosmological parameters, $\cpar = (H_0,\OM,\OL)$,
and $\spar$ denote possible extra parameters associated with
modelling evolution or other sources of systematic errors.
Our task is to find the posterior distribution for these parameters given
the data and some model, $M$.  Actually, we are ultimately interested
in the marginal distribution for $\OM$ and $\OL$, found by
marginalizing: $p(\OM,\OL|D,M) = \int dH_0 \int d\spar p(\cpar|D,M)$.  Bayes's
theorem gives the joint posterior distribution for $\cpar$ and $\spar$,
\be
p(\cpar,\spar|D,M) \propto p(\cpar|M) p(\spar|M) \like(\cpar,\spar).
\eqn(BT-CS)
\ee
The first factor is the prior for $\cpar$, which we will take to be
flat over the ranges shown in our plots (or flat in the logarithm for
$H_0$; see below).  The second factor is the prior for $\spar$ which we
assume is independent of $\cpar$; we discuss it further in the context
of specific models, below.  The last factor is the likelihood for
$\cpar$ and $\spar$, which we have abbreviated as  $\like(\cpar,\spar)
\equiv p(D|\cpar,\spar,M)$.  Rigorous calculation of this likelihood is
very complicated, requiring introduction and estimation of many
additional parameters, including parameters from the lightcurve model
and parameters for characteristics of the individual SNe (such as their
apparent and absolute magnitudes, redshifts, $K$-corrections, etc.).
With several simplifying assumptions, the final result is relatively
simple; it can be written as the product of independent Gaussians for
the redshifts and distance moduli of the SNe integrated over the
redshift uncertainty, so that
\be
\like(\cpar,\spar)
  \approx   
   \prod_i \int dz_i 
     \exp\left[-{[F(z_i) - \hat \mu_i]^2 \over 2s_i^2}\right]
     \exp\left[-{(z_i - \hat z_i)^2 \over 2w_i^2}\right]. 
\eqn(L-CS-3)
\ee
Here $\hat\mu_i$ is the best-fit distance modulus for SNe number $i$, $s_i$
is its uncertainty, $\hat z_i$ is the best-fit cosmological
redshift, and $w_i$ is its uncertainty (mostly due
to the source's peculiar velocity).  The function $F(z_i)$ gives
the true distance modulus for a SN Ia at redshift $z_i$; in the
absence of systematic or evolutionary terms, it is given by
$f(z_i)$ in \ceqn(eq:mu).  For the results
reported in P98, two complications appear in the likelihood.
First, the factors are not independent;
the use of common photometric calibration
data for groups of SNe Ia that are studied together introduces correlations.
P98 have reported a correlation matrix accounting for these, but
the correlations are very small and we have neglected them here.
In addition, one of the parameters defining the lightcurve model---the 
$\alpha$ parameter described in \S~2, above---appears explicitly in the 
P98 likelihood so that it can be estimated jointly with the cosmological
parameters.  This parameter would appear in the $\hat\mu_i$ estimates
in \ceqn(L-CS-3).  The data tabulated in P98 use the best-fit $\alpha$,
however, so we could not account for the uncertainty of $\alpha$ in
our analysis.  The close similarity between our contours in 
Figure~11$a$ and those presented in P98 argues that rigorous accounting
for the uncertainty in $\alpha$ plays only a minor role in the final
results.

As was done in R98 and P98, we approximate the $z_i$ integrals in
\ceqn(L-CS-3) by linearizing the $z_i$ dependence of $F(z_i)$ about
$\hat z_i$ and performing the resulting convolution of Gaussians
analytically.  The result is 
\be
\like(\cpar,\spar)
  \approx   
   \prod_i \exp\left[-{[\mu_i - \hat \mu_i]^2 \over 2\sigma_i^2}\right].
\eqn(L-CS)
\ee
where $\mu_i = F(\hat z_i)$ and 
\be
\sigma^2_i = s_i^2 + [F'(\hat z_i)]^2 w_i^2.
\eqn(sig-def)
\ee
The total variance $\sigma_i^2$ depends on $\cpar$ through
$F'(z)$.  But this dependence is weak in general, and $F'(z)$ is
actually independent of $\cpar$ at low redshift in the pure
cosmology model, with
\be
F'(z) = {5 \log e \over z}.
\eqn(Fp-def)
\ee
We follow the practice of R98 and simply use this formula for all
redshifts.  We use the same formula for models with systematic
error terms that introduce an
additional (weak) dependence on redshift and $\spar$; the redshift
uncertainties are negligibly small at high redshifts where such
dependences might become important, so the dependence of $\sigma_i^2$
on redshift is negligible.  It is possible to do the $z_i$ integrals in
\ceqn(L-CS-3)\ accurately using Gauss-Hermite quadrature.  We have done some
calculations this way and verified that the final inferences are
negligibly affected by the redshift integral approximations.

Equation \ceq(L-CS)\ is the starting point for the analyses reported
in the body of this work.  It is of a simple form: $-2$ times
the log-likelihood is of the form of a $\chi^2$ statistic.  
This is the same likelihood we would have written down had we
simply presumed at the outset that the reported $\hat\mu_i$ values
were equal to some underlying true values given by $F(\hat z_i)$
plus some added noise $n_i$;
\be
\hat \mu_i = F(\hat z_i) + n_i,
\eqn(mu-F-n)
\ee
where the probability distribution for the value of $n_i$ is
a zero-mean Gaussian with standard deviation $\sigma_i^2$.

\subsection{FRW Cosmology}

Presuming a FRW cosmology and no systematic errors, \ceqn(mu-F-n)\ can
be written,
\bea
\hat \mu_i 
  &=& f_i + n_i\nonumber\\
  &=& g_i - \eta + n_i,
\eqn(mu-g-eta)
\eea
where $f_i = f(\hat z_i)$ is the magnitude-redshift relation, which we
can separate into a part $g_i = g(\hat z_i)$ that depends implicitly only on 
$\OM$ and $\OL$, and the $H_0$ dependence is contained in $\eta$
(defined in \ceqn(eta-def)).
Define the quadratic form $Q$ according to
\be
Q = \sum_i {(\hat\mu_i - g_i + \eta)^2 \over \sigma_i^2}.
\eqn(Q-FRW)
\ee
This is the $\chi^2$ statistic used in R98; the joint likelihood for
$h$, $\OM$, and $\OL$ is simply proportional to $e^{-Q/2}$.  We 
can analytically marginalize over $h$ (or equivalently,
over $\eta$) to find the marginal likelihood for the density
parameters.  To do so, we must assign a prior for $h$.  We use the
standard noninformative ``reference'' prior for a positive scale
parameter, a prior flat in the logarithm and thus
scale-invariant (Jeffreys 1961; Jaynes 1968;
Yang and Berger 1997).  This corresponds to a prior that is flat in
$\eta$.  We bound this prior over some range $\Delta\eta$ (with limits
corresponding to $h=0.1$ and $h=1$, so $\Delta\eta = \ln[10]$).  The
prior range has negligible effect on all our results (so long as it
contains the peak of the likelihood) because the $H_0$ parameter is
common to all models, so the prior range cancels out of all probability
ratios.  Thus we could let it become infinite, but it is a good practice
in Bayesian calculations to always adopt proper (i.e., normalizable)
priors, especially if Bayes factors (ratios of normalization constants)
are of interest.

Using the log-flat prior, the marginal likelihood for the density
parameters is
\be
\like(\OM,\OL) = {1 \over \Delta\eta} \int d\eta e^{-Q/2}.
\eqn(L-ML-FRW-0)
\ee
To do the integral, complete the square in $Q$ as a function of $\eta$,
writing
\be
Q = {(\eta -\hat\eta)^2\over s^2} - q(\OM,\OL),
\eqn(Q-FRW-2)
\ee
where
\be
{1\over s^2} = \sum_i {1 \over \sigma_i^2},
\eqn(s-FRW)
\ee
\be
\hat\eta(\OM,\OL) = s^2 \sum_i {g_i - \hat\mu_i\over \sigma_i^2},
\eqn(ehat-FRW)
\ee
and the $(\OM,\OL)$-dependence is isolated in
\bea
q(\OM,\OL) 
  &=& - {\hat\eta^2\over s^2}
      + \sum_i {(\hat\mu_i - g_i)^2 \over \sigma_i^2}\nonumber\\
  &=& \sum_i {(\hat\mu_i - g_i + \hat\eta)^2 \over \sigma_i^2}.
\eqn(q-FRW)
\eea
The integral in \ceqn(L-ML-FRW-0)\ is thus simply an integral over a
Gaussian in $\eta$ located at $\hat\eta$ with standard deviation $s$; $\hat\eta$ is the
best-fit (most probable) value of $\eta$ given $\OM$ and $\OL$, and $s$
is its conditional uncertainty.  As long as $\eta$ is inside the prior
range and $s\ll \Delta\eta$, the value of this integral is well
approximated by $s\sqrt{2\pi}$, so that
\be
\like(\OM,\OL) = {s \sqrt{2\pi}\over \Delta\eta} 
     e^{-q/2}.
\eqn(L-ML-FRW)
\ee
This is the marginal likelihood one would use to infer the density
parameters in the absence of any systematic error terms.  Note
from \ceqn(q-FRW)\ that the quadratic form is just what one would
obtain by calculating the ``profile likelihood'' for the density
parameters (the likelihood maximized over the nuisance parameters,
a frequentist method sometimes used to approximately treat
nuisance parameters).
Since the uncertainty $s$ is independent of $\OM$ and $\OL$,
it follows from \ceqn(L-ML-FRW)\ that the marginal likelihood
is proportional to the profile likelihood in this problem.

It is also possible to do the calculation analytically using a
flat prior for $h$, spanning a prior range $\Delta h$.  The
corresponding prior for $\eta$ is exponential;
\be
p(\eta|M) = {10^5 c_2 \over 2a\Delta h} e^{\eta/2a},
\eqn(h-flat)
\ee
where $a = 2.5\log e$, a constant known as Pogson's ratio (Pogson 1856).
The product of the likelihood and the prior can still be written
as $e^{-Q/2}$ with $Q$ quadratic in $\eta$; but there is an additional
linear term in $Q$ from the prior.  Completing the square duplicates \ceqn(Q-FRW-2),
but with $\hat\eta$ replaced with
\be
\hat\eta = s^2 \left[-{1 \over 4a} + 
      \sum_i {g_i - \hat\mu_i\over \sigma_i^2}\right].
\eqn(ehat-FRW-flat)
\ee
The marginal likelihood for $\OM$ and $\OL$ also has a different
factor out front; it is given by
\be
\like(\OM,\OL) = {10^5 c_2 s \sqrt{2\pi}\over 2 a\Delta h} 
     e^{-q/2}.
\eqn(L-ML-FRW-flat)
\ee
We present this result for reference only; we use the scale-invariant 
prior in the body of this work and in the remainder of this Appendix.
We have compared calculations with flat and log-flat priors for
some models; the resulting marginal likelihoods are negligibly
different.  Note that \ceqn(L-ML-FRW-flat)\ is not proportional to
the profile likelihood; the proportionality is a special property
of the scale-invariant prior.

\subsection{Systematic Error in $H_0$}

Among the lightcurve model parameters, regardless of the method, is the
fiducial absolute magnitude for SNe Ia, $M_0$.  To obtain definite values for the
distance moduli, $M_0$ must be estimated or at least arbitrarily
specified.  Let $\hat M_0$ denote the value used to calculate the
tabulated $\hat\mu_i$ estimates.  We can write the true value as
\be
M_0 = \hat M_0 + \delta,
\eqn(M0-d)
\ee
where $\delta$ is an uncertain error in our estimate.  Since the $\hat\mu_i$
estimates are calculated using $\hat M_0$, they will have an
additive error equal to $\delta$ that is systematic (the same for
every SNe Ia).  To account for this, \ceqn(mu-g-eta)\ must be
replaced by
\be
\hat \mu_i = g_i - \eta + \delta + n_i.
\eqn(mu-g-eta-d)
\ee
Note here the degeneracy between $\eta$ and $\delta$; since they play
identical roles (up to a sign) in the model for the distance moduli,
they cannot be individually constrained using only magnitude/redshift data;
additional information setting a distance scale to at least one SN is
required.  Only the quantity $\gamma = \delta - \eta$ can be inferred
from the basic data.  

P98 arbitrarily specify $\hat M_0$, so there is no useful information
about $\delta$ that can break the degeneracy between $\delta$ and
$\eta$.  Recognizing this, they simply forgo any attempt to
infer the Hubble constant.  Their
analysis amounts to replacing $\eta$ and $\delta$ with $\gamma$ and
marginalizing over $\gamma$ with a flat prior; the resulting marginal
likelihood for the density parameters is of the same form as
\ceqn(L-ML-FRW), though with an arbitrarily large prior range
for $\gamma$ (which can be ignored since it is common to all models
being compared).  This is the likelihood we used for the analyses of
LBL data (and simulated data) described in \S~4 when we assume
no evolutionary effects are present.

R98 use Cepheid distances for three SNe Ia to estimate $M_0$ for use
with the MLCS and TF methods.  We can consider this extra data to
provide a prior distribution for $\delta$; this prior breaks the
degeneracy between $\eta$ and $\delta$ in the analysis.  R98 report a
10\% uncertainty in the Cepheid distance scale for SNe Ia,
corresponding to 0.21 magnitude uncertainty in distance moduli.  We
accordingly adopt a Gaussian prior for $\delta$ with zero mean and
standard deviation $d=0.21$, so that
\be
p(\delta) = {1 \over d\sqrt{2\pi}} e^{-\delta^2/2d^2}.
\eqn(p-delta)
\ee
We can calculate the likelihood for the cosmological parameters
by multiplying the joint likelihood for them and $\delta$ by
this prior, and integrating over $\delta$, as follows.

The quadratic form in the exponential resulting from multiplying
this prior by the likelihood resulting from \ceqn(mu-g-eta-d)\ is,
\bea
Q 
  &=& {\delta^2 \over d^2} +
        \sum_i {(\hat\mu_i - g_i + \eta - \delta)^2 \over \sigma_i^2}\nonumber\\
  &=& {(\delta -\hat\delta)^2\over s^2} - q(\OM,\OL),
\eqn(Q-delta)
\eea
where
\be
{1\over s^2} = {1 \over d^2} + \sum_i {1 \over \sigma_i^2},
\eqn(s-delta)
\ee
\be
\hat\delta(\eta,\OM,\OL) = s^2 \sum_i {\hat\mu_i - g_i + \eta\over \sigma_i^2},
\eqn(dhat-delta)
\ee
and the $(\eta,\OM,\OL)$-dependence is isolated in
\bea
q(\OM,\OL) 
  &=& - {\hat\delta^2\over s^2}
      + \sum_i {(\hat\mu_i - g_i + \eta)^2 \over \sigma_i^2}\nonumber\\
  &=& {\hat\delta^2\over d^2} + 
      \sum_i {(\hat\mu_i - g_i + \eta - \hat\delta)^2 \over \sigma_i^2}.
\eqn(q-delta)
\eea
As with $\eta$ in the previous subsection, the integral over $\delta$ is
a simple Gaussian integral, equal to $s\sqrt{2\pi}$.  Thus the marginal
likelihood for the cosmology parameters is
\be
\like(\eta,\OM,\OL) = {s \over d} e^{-q/2}.
\eqn(L-hML-delta)
\ee
This is the likelihood used for the analyses of the MLCS and TF data using
the default model, as reported in \S~4.1.

\subsection{Systematic Error From Evolution}

The simplest model we considered with a redshift-dependent systematic
or evolutionary component is Model~I, which adds a shift of size $\eps$
to the distance moduli of the high redshift SNe Ia.  For this model,
\be
\hat \mu_i = \left\{ \begin{array}{ll}
      f_i + \delta + n_i & \mbox{if $z_i < z_c$}\\
      f_i + \delta + \eps + n_i & \mbox{if $z_i \ge z_c$}
   \end{array}\right.,
\eqn(mu-eps)
\ee
with $z_c = 0.15$.  We seek the marginal likelihood for the cosmological
parameters, requiring us to introduce priors for $\delta$ and $\eps$
and marginalize over them.

The prior for $\delta$ is given by \ceqn(p-delta), and the prior
for $\eps$ is similarly a zero-mean Gaussian, but with a different
standard deviation, $e$;
\be
p(\eps) = {1 \over e\sqrt{2\pi}} \exp\left[-{\eps^2\over 2e^2}\right].
\eqn(p-eps)
\ee
The quadratic form associated with the product of these priors and
the likelihood function is
\be
Q = {\delta^2 \over d^2} + {\eps^2 \over e^2} +
    \sum_{z_i<z_c} {(\hat\mu_i - f_i - \delta)^2 \over \sigma_i^2} +
    \sum_{z_i\ge z_c} {(\hat\mu_i - f_i - \delta - \eps)^2 \over \sigma_i^2}.
\eqn(Q-de)
\ee
To marginalize over $\eps$, we complete the square in $\eps$ by introducing the
$\eps$ uncertainty $t$, given by
\be
{1\over t^2} = {1 \over e^2} + \sum_{z_i\ge z_c} {1 \over \sigma_i^2},
\eqn(t-de)
\ee
and the conditional best-fit value of $\eps$,
\be
\hat\eps(\delta,\cpar) = t^2 \sum_{z_i\ge z_c} 
    {\hat\mu_i - f_i -\delta\over \sigma_i^2}.
\eqn(ehat-de)
\ee
After completing the square and integrating the resulting Gaussian
dependence on $\eps$, we find that
\be
p(\delta)\like(\delta,\cpar) = {t \over ed\sqrt{2\pi}} e^{-q/2},
\eqn(like-de-1)
\ee
where
\be
q = - {\hat\eps^2 \over t^2} + {\delta^2 \over d^2} + 
   \sum_i {(\hat\mu_i - f_i - \delta)^2 \over \sigma_i^2}.
\eqn(q-de)
\ee
Note that the sum is over all SNe, and that $\delta$ appears
in $\hat\eps$.  Completing the square in $\delta$ lets us
identify the $\delta$ uncertainty, $s$, given by
\be
{1\over s^2} = {1 \over d^2} + \sum_i {1 \over \sigma_i^2}
   - {t^2 \over v^2},
\eqn(s-de)
\ee
and the conditional estimate for $\delta$,
\be
\hat\delta(\cpar) = s^2 \left[ -{t^2 \over v^2}F +
       \sum_i {\hat\mu_i - f_i\over \sigma_i^2} \right],
\eqn(dhat-de)
\ee
where in these equations we have defined $v$ and $F$ according to
\be
{1 \over v} = \sum_{z_i\ge z_c} {1 \over \sigma_i^2},
\eqn(v-def)
\ee
and
\be
F = \sum_{z_i\ge z_c} {\hat\mu_i - f_i\over \sigma_i^2}.
\eqn(F-def)
\ee
Using these, we can rewrite $q$ as
\be
q = {(\delta - \hat\delta)^2 \over s^2} + q'(\cpar),
\eqn(q-de-2)
\ee
where the dependence on the cosmological parameters is in
\be
q'(\cpar) = -{\hat\delta^2\over s^2} - t^2 F^2 + 
   \sum_i {(\hat\mu_i - f_i)^2 \over \sigma_i^2}.
\eqn(qp-de)
\ee
After integrating over the Gaussian dependence on $\delta$,
the marginal likelihood is
\be
\like(\cpar) = {ts\over ed} e^{-q'/2}.
\eqn(like-dem)
\ee
This is the likelihood used for analyses of the MLCS and
TF data based on Model~I.

For Model~II, used to model the SF data, the estimated distance moduli
are given by
\be
\hat \mu_i = g_i + \gamma + \beta h_i + n_i,
\eqn(mu-beta)
\ee
where as before $\gamma = \delta - \eta$, and $h_i = \ln(1+\hat z_i)$.
We will marginalize over $\gamma$ and $\beta$, using a
flat prior for $\gamma$ and a zero-mean Gaussian prior for $\beta$
with standard deviation $b$.

As already noted, the $\gamma$ marginalization is similar to the
$\eta$ marginalization already treated above.  The result is
\be
p(\beta) \like(\OM,\OL) = {s \over \Delta\gamma b} 
     e^{-q/2},
\eqn(L-bML-beta)
\ee
where $s$ is given by \ceqn(s-FRW),
\be
\hat\gamma(\beta,\OM,\OL) = 
   s^2 \sum_i {\hat\mu_i - g_i - \beta h_i \over \sigma_i^2},
\eqn(ghat-beta)
\ee
and the $(\beta,\OM,\OL)$-dependence is isolated in
\bea
q(\beta,\OM,\OL) 
  &=& - {\hat\gamma^2\over s^2}
      + \sum_i {(\hat\mu_i - g_i - \beta h_i)^2 \over \sigma_i^2}\nonumber\\
  &=& \sum_i {(\hat\mu_i - g_i - \beta h_i - \hat\gamma)^2 \over \sigma_i^2}.
\eqn(q-beta)
\eea
We assume that the prior range for $\gamma$, $\Delta\gamma$, contains the peak of
the Gaussian.  Since this range is common to all models for this
data and thus cancels in all calculations, we do not need to
specify it more precisely, and we simply drop it from subsequent
calculations.

Note that $\hat\gamma$ depends on $\beta$; we can isolate this dependence
by writing
\be
\hat\gamma = s^2 H - \beta s^2 G,
\eqn(ghat-beta-2)
\ee
where
\be
H = \sum_i {\hat\mu_i - g_i \over \sigma_i^2},
\eqn(H-def)
\ee
and
\be
G = \sum_i {h_i \over \sigma_i^2}.
\eqn(G-def)
\ee
This helps us to do the remaining marginalization over $\beta$. 
We now complete the square in
$\beta$, identifying the $\beta$ uncertainty, $\tau$, given by
\be
{1 \over \tau^2} = {1 \over b^2} - s^2 G^2 + \sum_i {h_i^2 \over \sigma_i^2},
\eqn(tau-beta)
\ee
and the conditional best-fit $\beta$,
\be
\hat\beta = \tau^2\left[ - s^2 GH + 
      \sum_i {h_i(\hat\mu_i - g_i) \over \sigma_i^2}\right].
\eqn(bhat-beta)
\ee
Integrating over the Gaussian dependence on $\beta$ gives
a factor of $\tau\sqrt{2\pi}$, and the final likelihood for 
the density parameters is
\be
\like(\OM,\OL) = {s \tau \sqrt{2\pi}\over b} e^{-q'/2},
\eqn(like-beta)
\ee
where
\be
q' = {\hat\beta^2 \over b^2} +
   \sum_i {(\hat\mu_i - g_i - \hat\beta h_i - s^2 H)^2 \over \sigma_i^2}.
\eqn(qp-beta)
\ee
This is the likelihood used for the calculations with Model~II in
\S~4.

\clearpage

\clearpage

\figcaption{Joint credible regions for $\Omega_\Lambda$ versus
$\Omega_m$ for the data set of P98.  The 68.3\%, 95.4\% and 99.7\%
confidence level contours are shown for ({\em a}) the data as published
and ({\em b}) after introducing a systematic offset of $-0.1$
magnitudes to the high redshift $(z > 0.15)$ sample.}

\figcaption{Histogram of the corrections to the absolute magnitudes of
the observed SNe Ia, in magnitudes, deduced from the MLCS, TF, and SF
methods.}

\figcaption{({\em a}) $\mu_{MLCS}$ versus $\mu_{TF}$ for the 10 high
redshift SNe Ia of R98.  ({\em b}) $\mu_{MLCS}$ versus $\mu_{SF}$ for
the 14 low reshift SNe Ia from the Cal\'an Tololo survey that are used
in both R98 and P98.  The dashed lines are straight line fits to the
data where the slope of the line is fixed to 1.}

\figcaption{Scatterplots comparing SNe Ia properties inferred using the MLCS and
TF methods for the R98 sample of SNe Ia.  Compared are ({\em a}) the host
galaxy extinction, $A$, for the 37 well measured SNe Ia; ({\em b}) the 
correction to the absolute magnitude $\Delta$ for the 37 well measured
SNe Ia; and ({\em c}) the peak apparent magnitude, $m$, of the 10 well
measured SNe Ia at high redshift.  The errors on $A$ and $\Delta$ can
be estimated to be of the order of 0.1 magnitudes.  The dashed lines
each have a slope of 1.}

\figcaption{The difference between the distance moduli inferred using the
MLCS and TF lightcurve fitting methods, $\Delta\mu = \mu_{MLCS} -
\mu_{TF}$, as a function of redshift $z$.  The errors on the data
points are described in the text.}

\figcaption{The difference between the distance moduli,  $\Delta\mu =
\mu_{MLCS} - \mu_{TF}$, is plotted versus an estimate of the absolute
magnitude, $M_B^{AV}$, for SNe Ia at low redshift ($z < 0.15$) in the
left hand plot and high redshift ($z > 0.15$) in the right hand plot.
The dashed line in the right hand plot is the result of a least squares
fit to the data which gives a slope of $-0.6\pm0.15$.}

\figcaption{The 68.3\% joint credible regions plotted
separately for intrinsically dim and intrinsically bright SNe Ia for
the ({\em a}) MLCS, ({\em b}) TF, and ({\em c}) SF analysis methods.
The contours for the full data set (all $M_B^{AV}$) are
shown as dashed curves.}

\figcaption{Marginal posterior distribution for $h$ using the MLCS and TF 
estimated distance moduli.  The labeled solid curves show results
using our reference model incorporating
systematic uncertainties in the Cepheid
distances for the SNe Ia used to set the absolute magnitude
scale of SNe Ia.  The dotted curves show results using Model~I,
with the prior uncertainty for the high-$z$ offset $e=0.2$ mag.}

\figcaption{The 68.3\%, 95.4\% and 99.7\% joint credible regions
for $\OM$ and $\OL$ based on our reference model, using
distance moduli calculated with the (a) MLCS and (b) TF lightcurve
fitting methods.}

\figcaption{The 68.3\%, 95.4\% and 99.7\% joint credible regions for for
$\OM$ and $\OL$ based on Model I. Panels (a) and (b) are for MLCS and
TF, respectively, for $e=0.1$, and (c) and (d) are for MLCS and TF,
respectively, with $e=0.2$.}

\figcaption{The 68.3\%, 95.4\% and 99.7\% joint credible regions
for Model II applied to distance moduli calculated with the SF lightcurve
fitting method. Panel (a) is for no evolution, and 
shown for reference. Panels (b) and (c) are for $b=0.25$
and $b=0.5$, respectively.}

\figcaption{The solid line shows the Bayes factor for $\OL\neq 0$
versus  $\OL=0$ as a function of the parameter $b$ of Model II,
and the dashed line shows the Bayes factor for Model II versus
the zero-evolution reference model.}

\figcaption{Marginal posterior distributions for $\OM$ (and,
equivalently, for $\OL = 1 - \OM$) presuming a flat cosmology, using
data from the MLCS (top), TF (middle), and SF (bottom) methods.
Results are shown presuming no evolution (solid curves), allowing a
small amount of evolution (short-dashed), and allowing a larger amount
of evolution (long-dashed).}

\figcaption{Comparison of $\mu$ in cosmological models with and without
evolution. The thick solid line is for the best-fit cosmology for
SF presuming no evolution $(\OM,\OL)=(0.75,1.34)$. The dotted curve
is for $(\OM,\OL)=(1,0)$ without evolution; the dashed curve is
$\beta\ln(1+z)$ with the best-fit value $\beta=0.83$ for this
cosmology. The thin solid line is the sum, and depicts
the best-fit presuming $(\OM,\OL)=(1,0)$ with evolution included.
The bold, long-dashed line is for $(\OM,\OL)=(0.3,0.7)$
without evolution.}

\figcaption{Results of analyzing simulated data with
$(\OM,\OL,\beta)=(1,0,0.5)$. Panels (a) and (b) are for analyses
presuming no evolution using data sets with 38 and 186 high-redshift
($0.3\leq z\leq 1.5$) sources, respectively; both data sets had 14
low-redshift sources. Panel (c) repeats the analysis of the larger data
set with evolution included in the model.  The crosses indicate
the best-fit parameter values from the analyses; the dots
indicate the true values used to generate the data.}

\clearpage

\begin{deluxetable}{lrr}
\tablecolumns{3}
\tablewidth{0pc}
\tablecaption{Dispersion of the data, in magnitudes, from 
the best fit cosmology for
low $z$ ($z < 0.15$) and high $z$ ($z > 0.15$).}
\tablehead{
\colhead{Fitting Method}    &   \colhead{With Corrections}   &
\colhead{Without Corrections} } 
\startdata
MLCS & & \\
\hfill low z & $0.18 \pm 0.02$ & $0.33 \pm 0.05$\\
\hfill high z& $0.22 \pm 0.05$ & $0.20 \pm 0.04$\\
\hline
TF & & \\
\hfill low z & $0.20 \pm 0.03$ & $0.33 \pm 0.05$\\
\hfill high z& $0.17 \pm 0.04$  &$0.20 \pm 0.05$ \\
\hline
SF & & \\
\hfill low z & $0.18 \pm 0.03$ & $0.18 \pm 0.03$ \\
\hfill high z& $0.30 \pm 0.03$ & $ 0.30 \pm 0.03$ \\
\enddata
\end{deluxetable}

\begin{deluxetable}{lll}
\tablecolumns{3}
\tablewidth{0pc}
\tablecaption{Interpretation of Bayes Factors}
\tablehead{
\colhead{$\ln(B_{ij})$}    &   \colhead{$B_{ij}$}   &
\colhead{Strength of evidence for $H_i$ over $H_j$} } 
\startdata
0 to 1 & 1 to 3 & Not worth more than a bare mention\\
1 to 3 & 3 to 20 & Positive\\
3 to 5 & 20 to 150 & Strong\\
$>5$ & $>150$ & Very Strong\\
\enddata
\end{deluxetable}

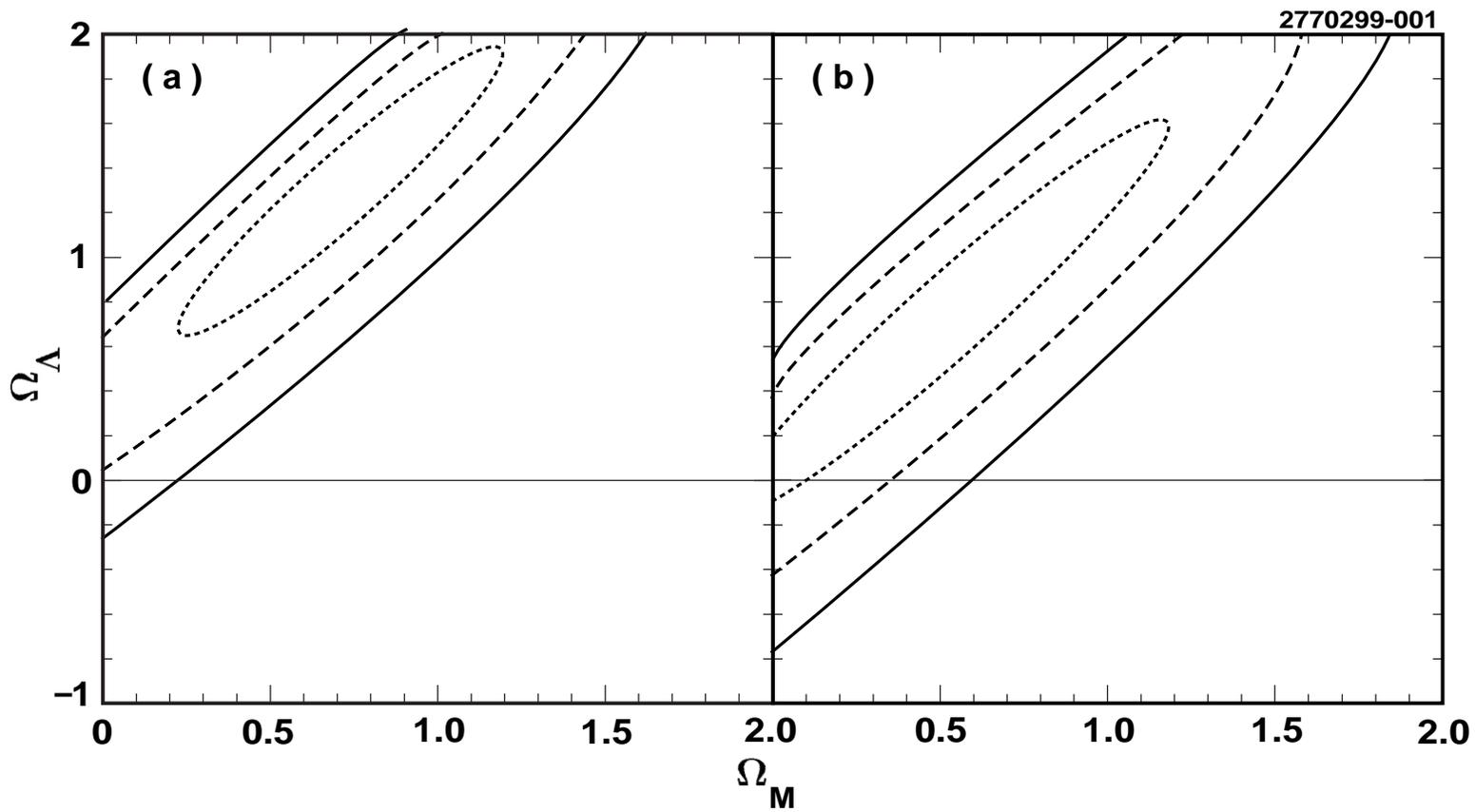
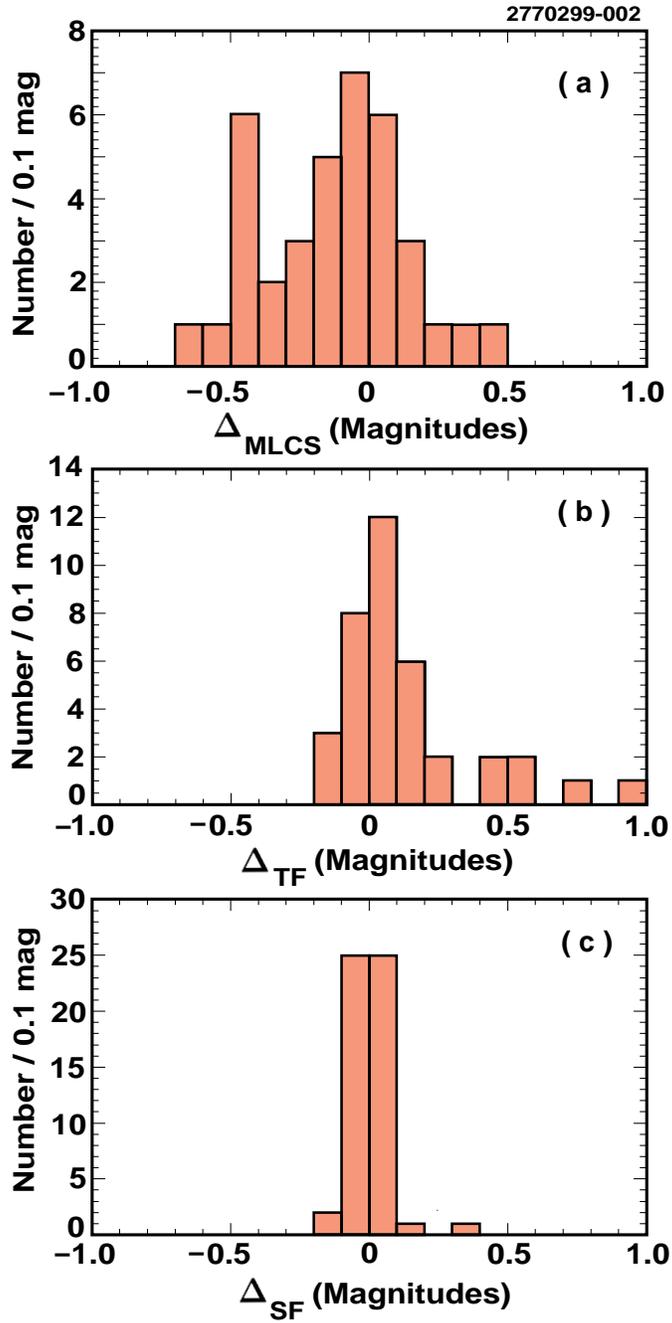
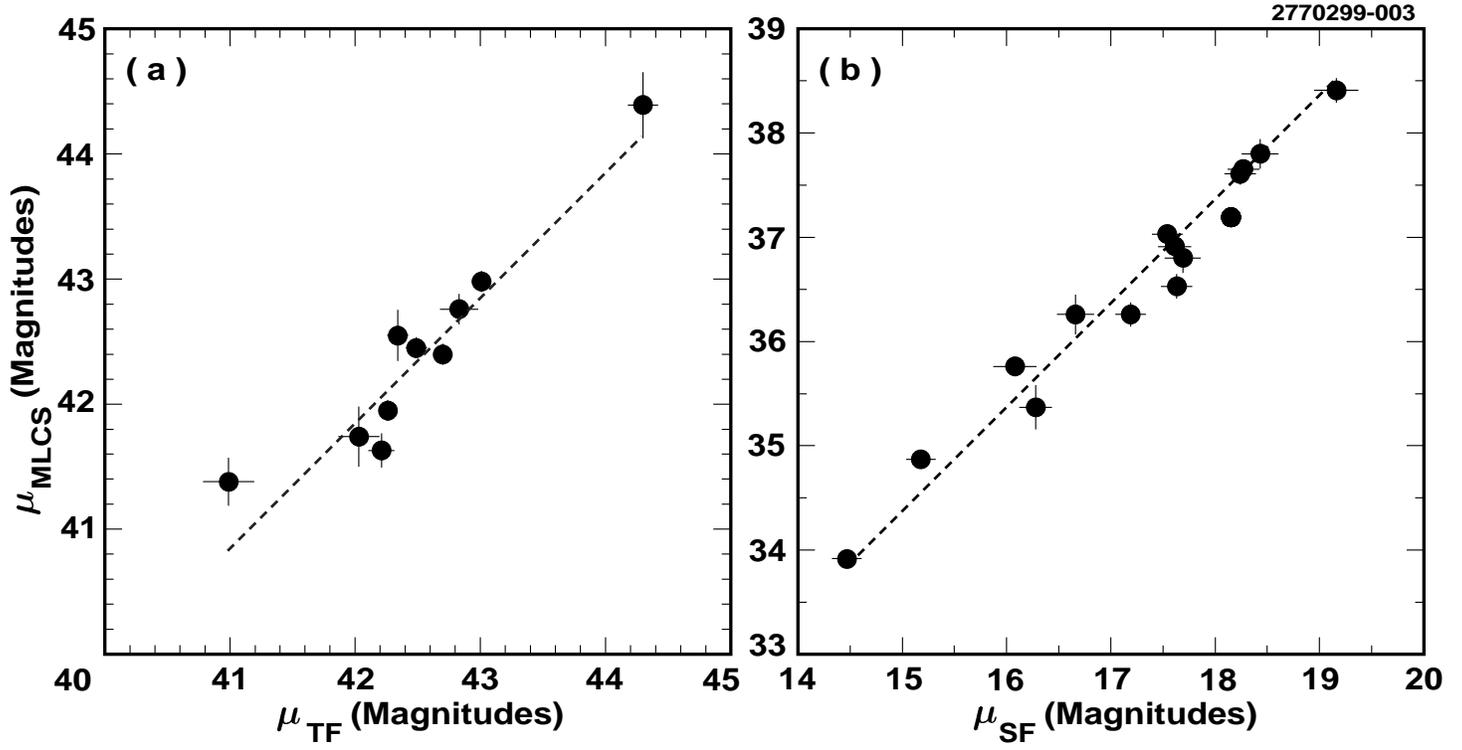
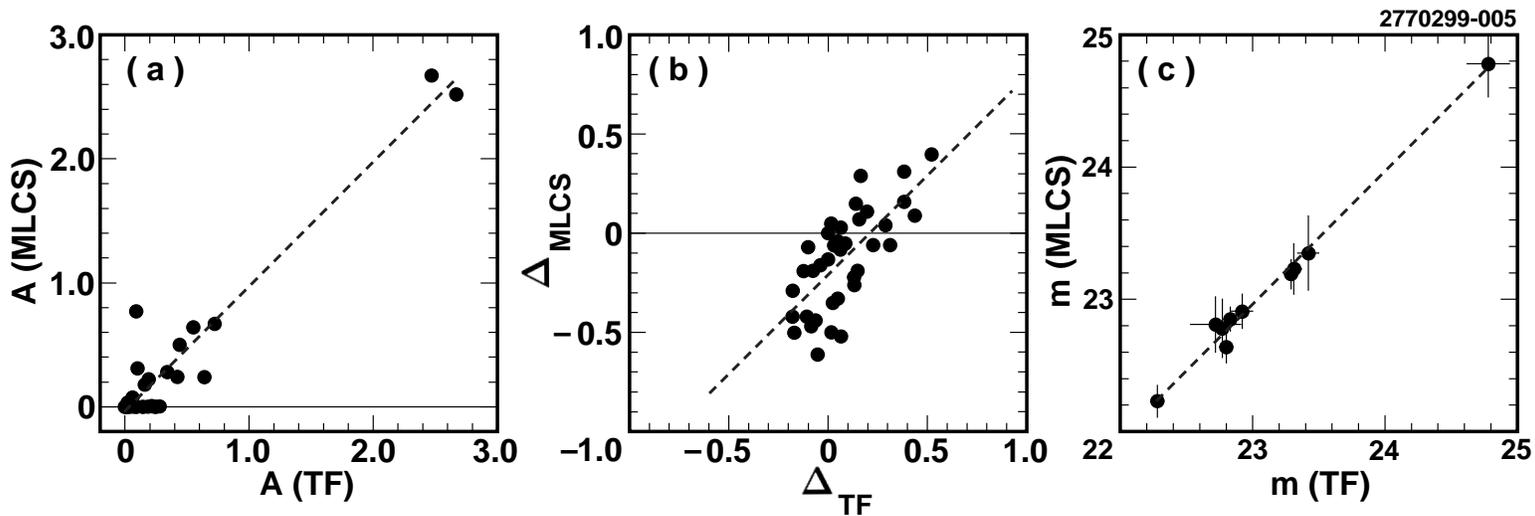
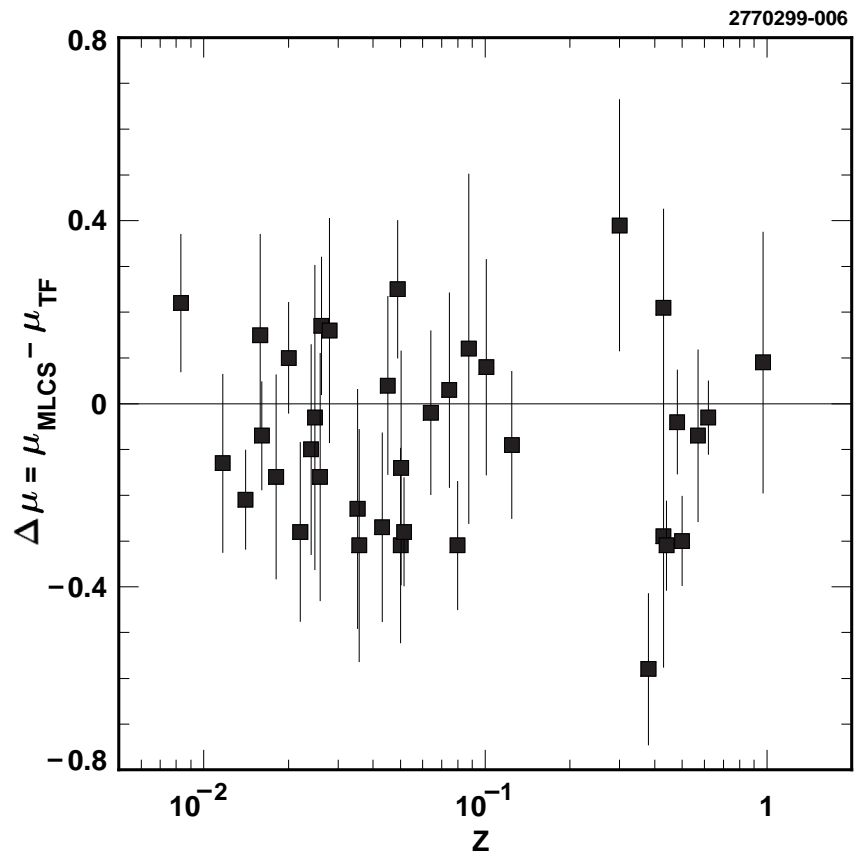
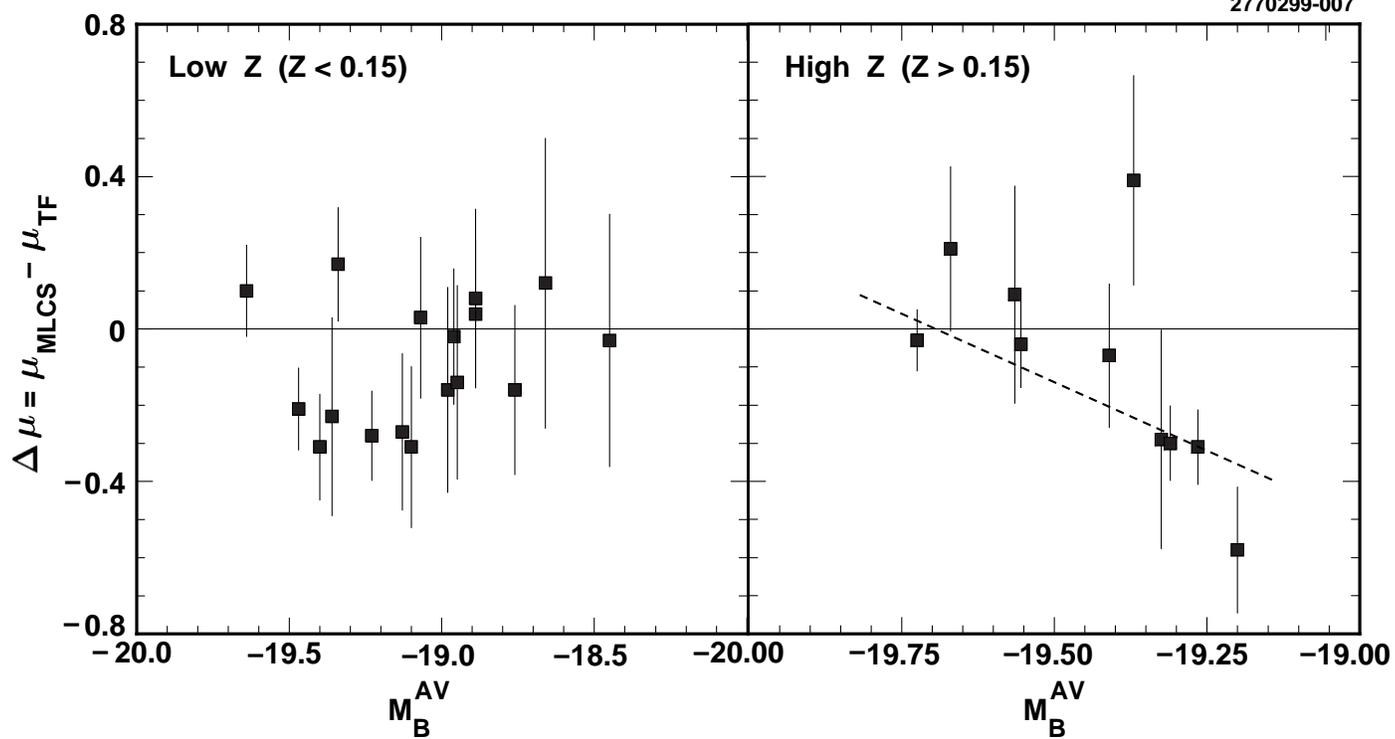
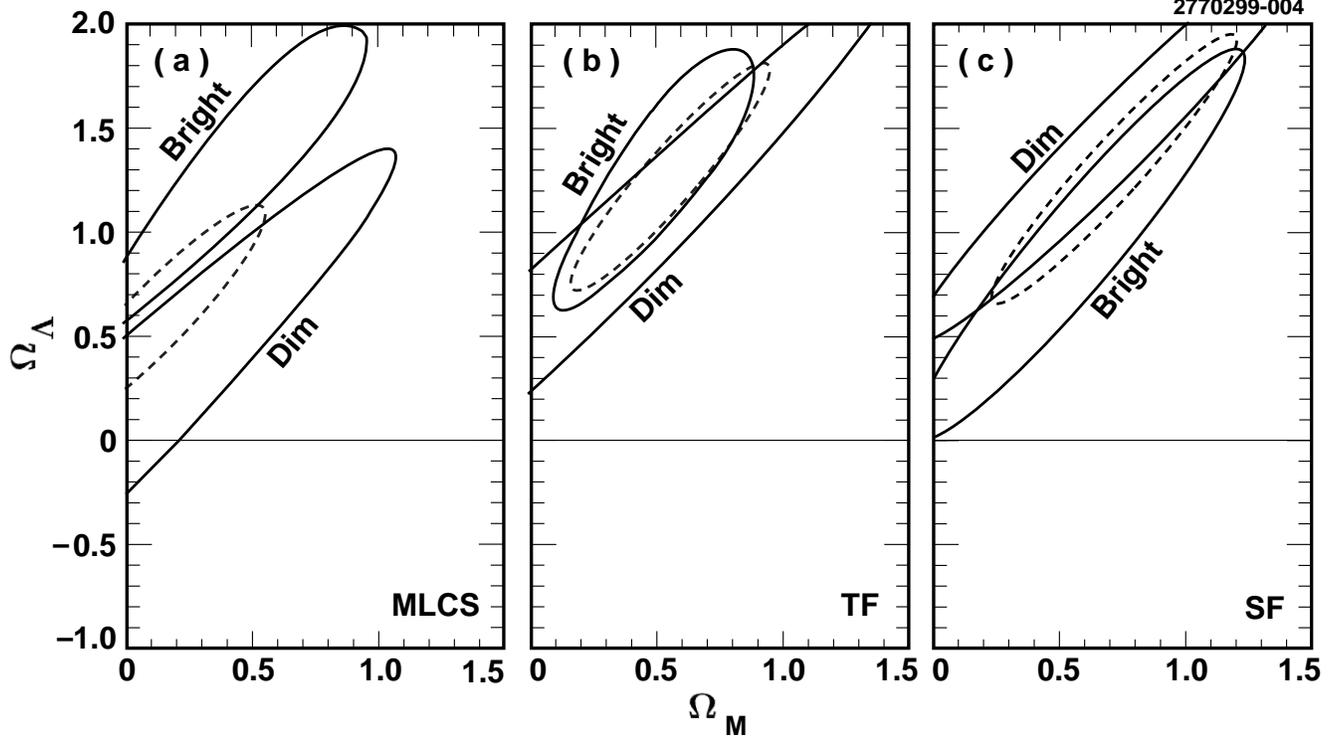
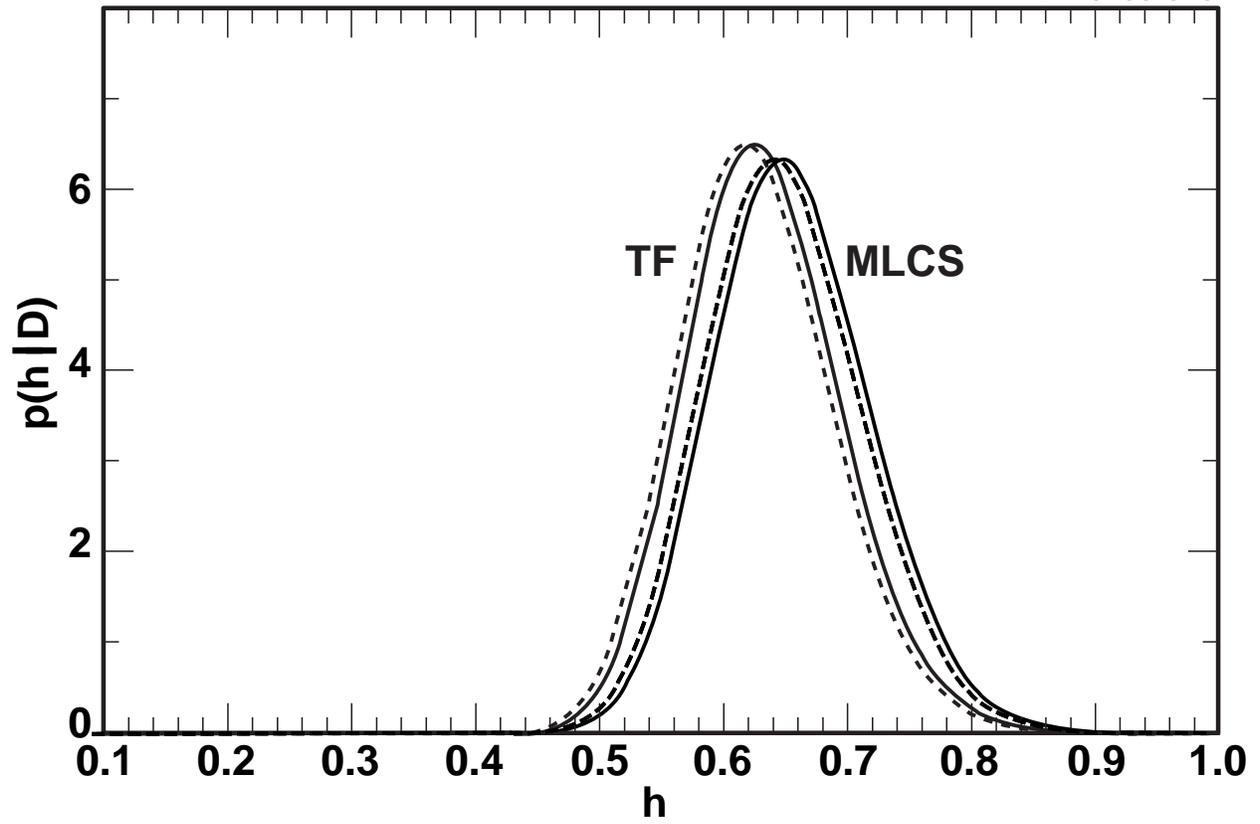
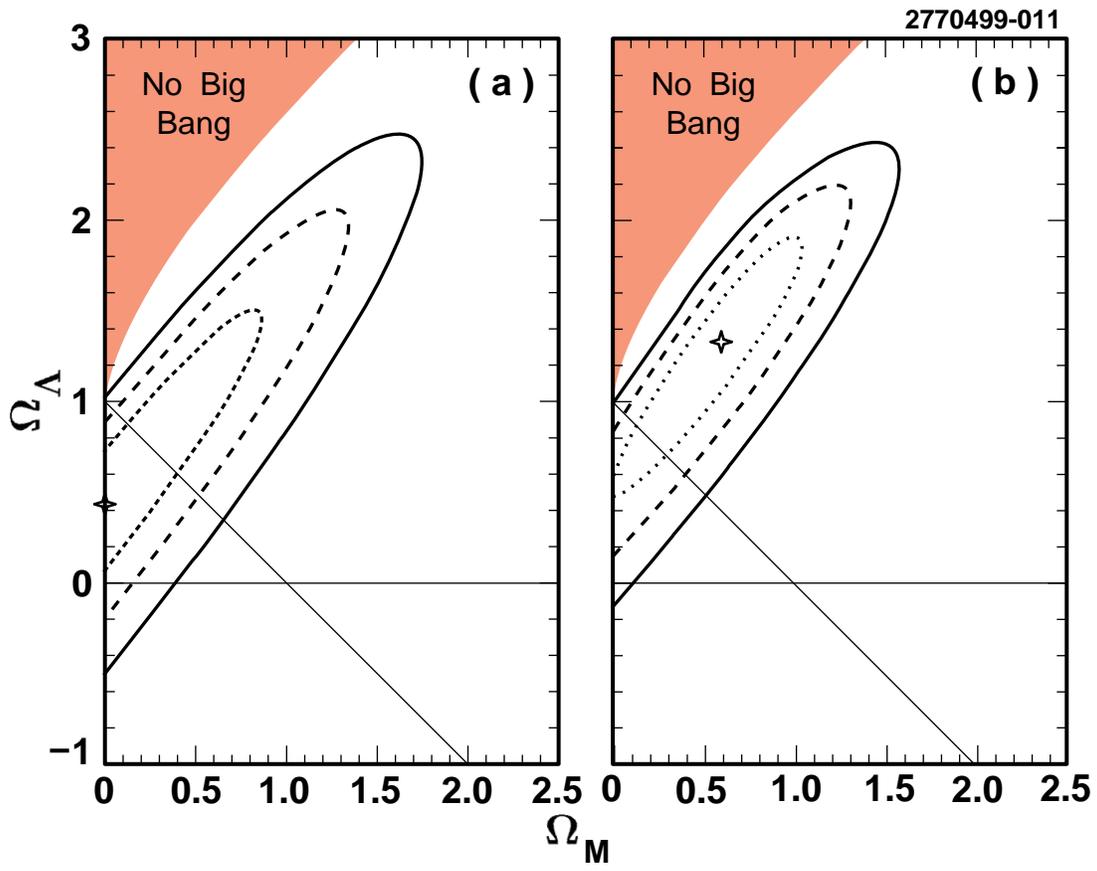
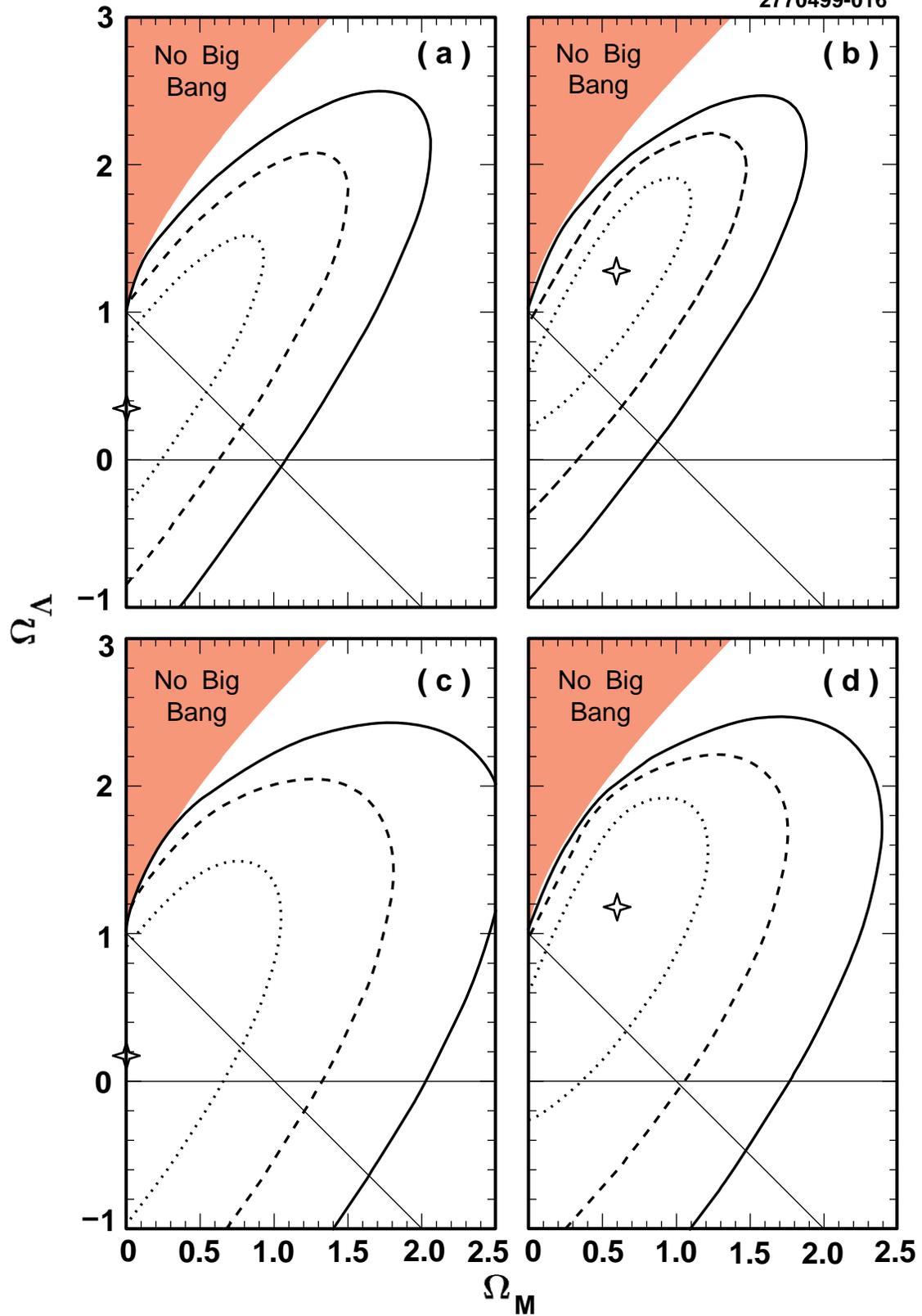
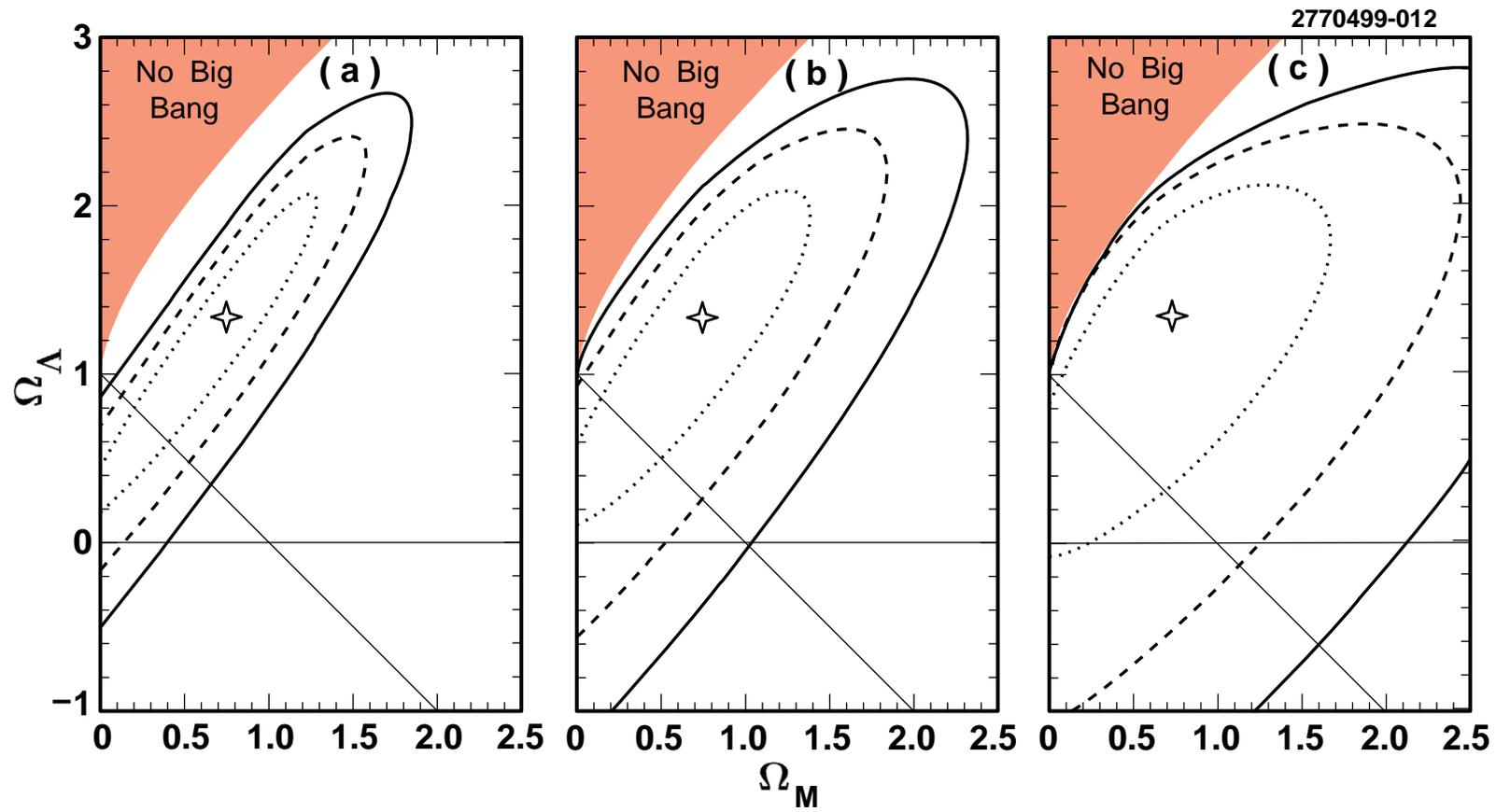
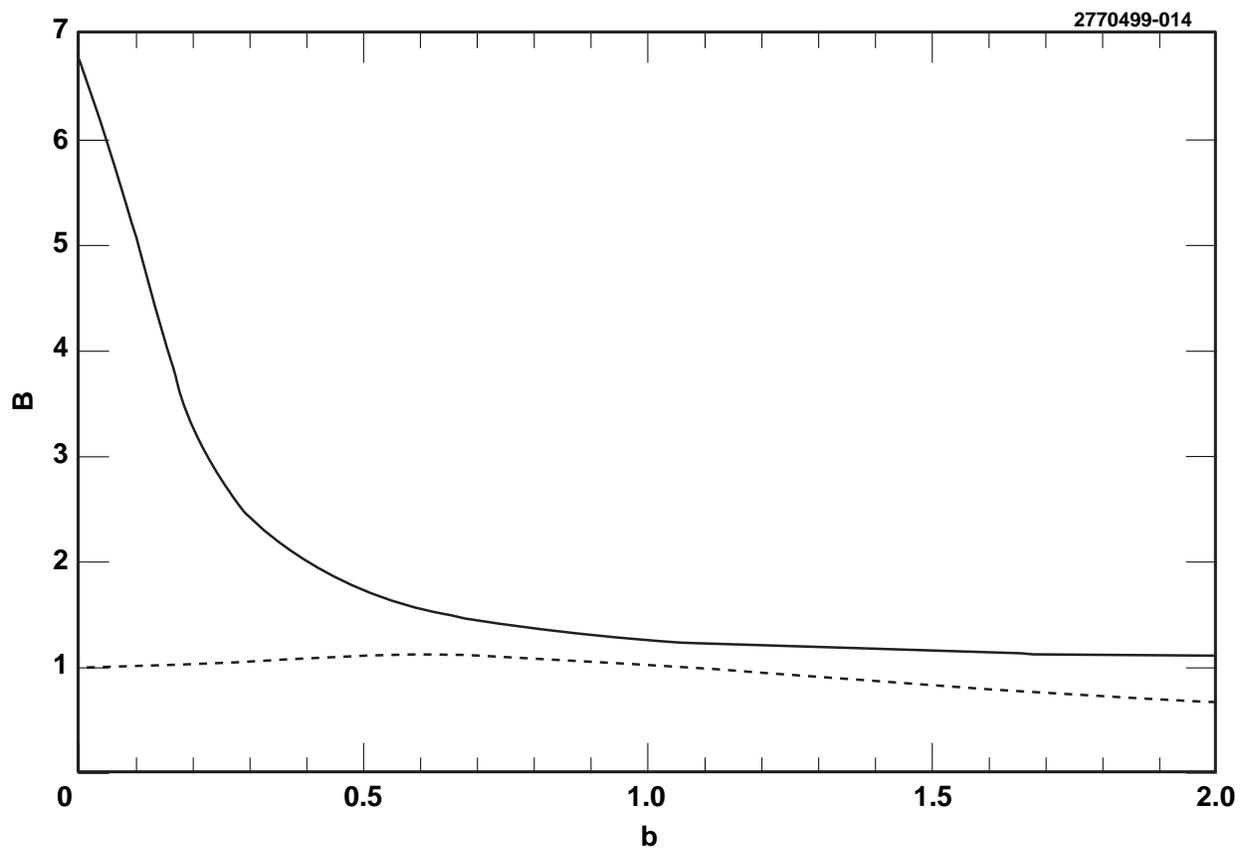
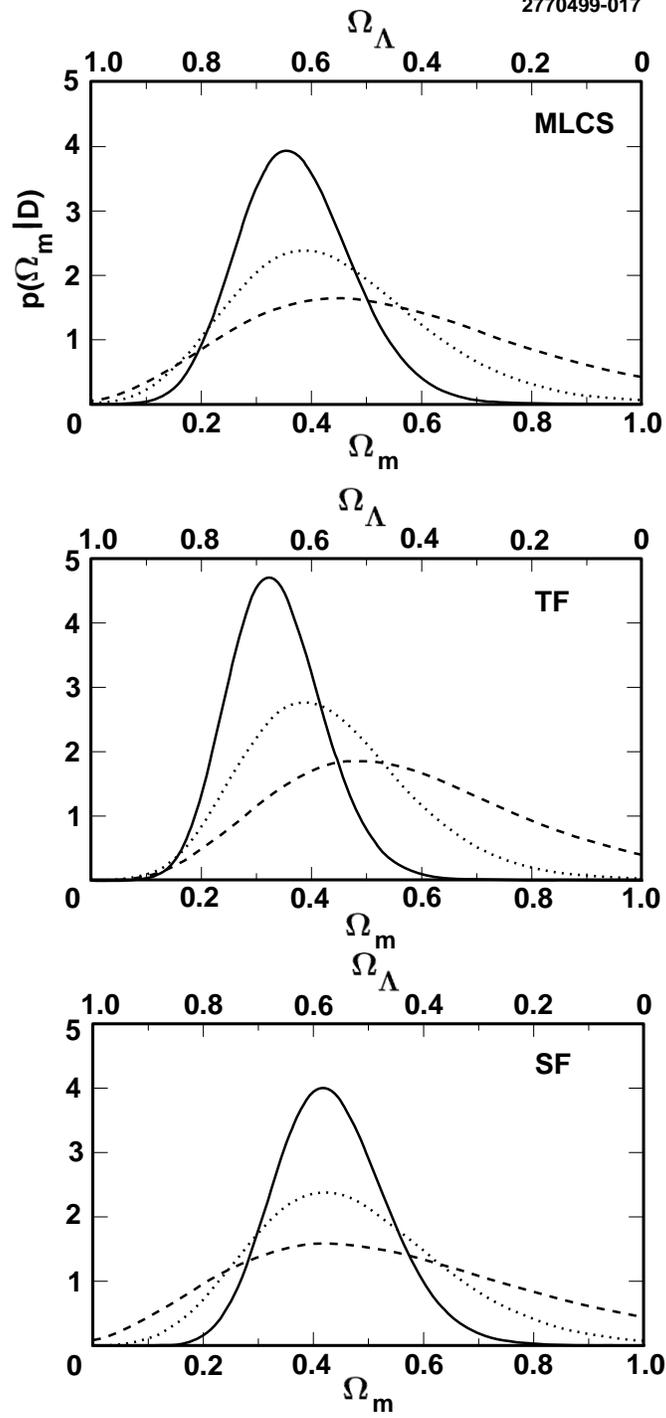
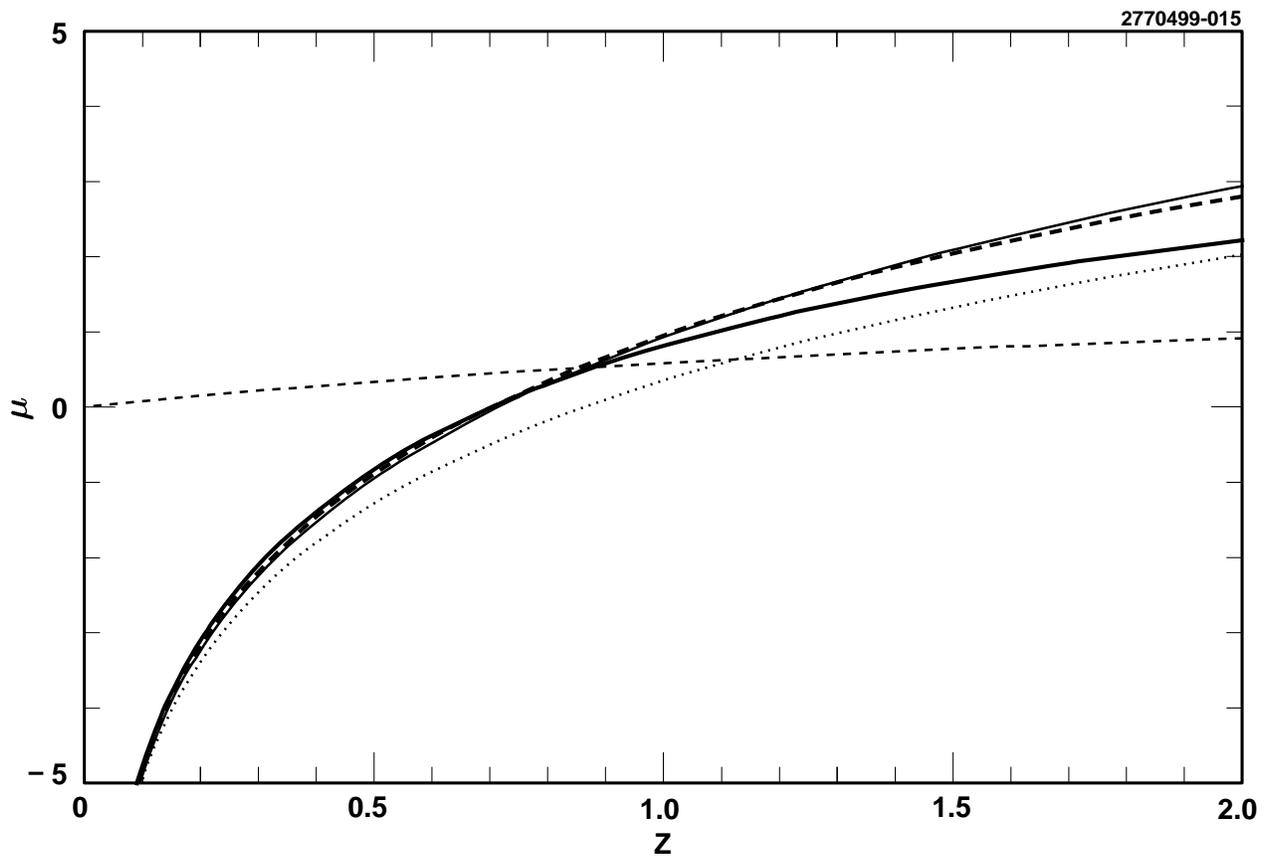
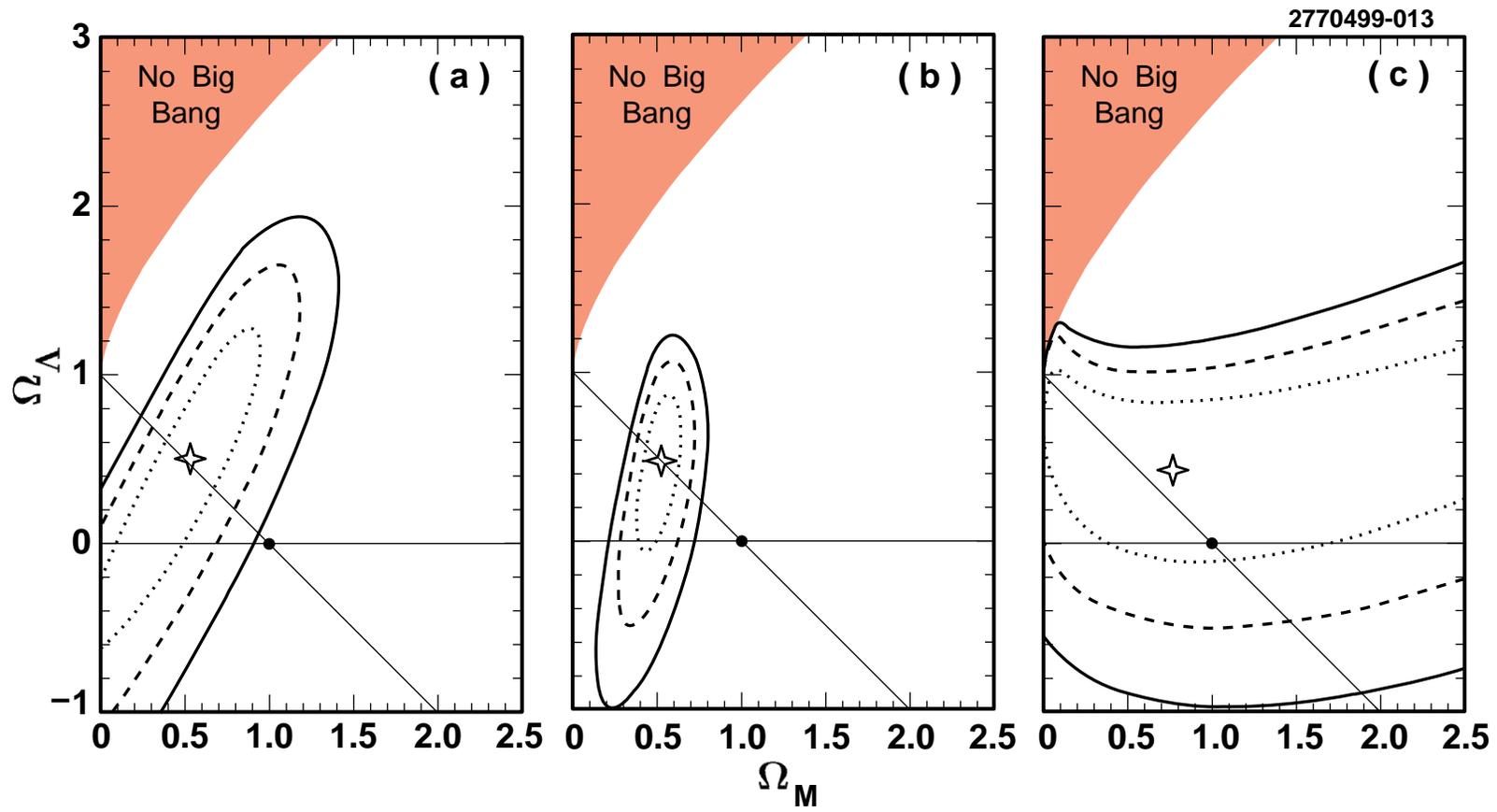
\end{document}